\documentclass[two column]{aastex63}   	
\usepackage{graphicx}				
\usepackage{amssymb, rotating}

\begin{document}

\title{Observing Carbon \& Oxygen Carriers in Protoplanetary Disks at Mid-infrared Wavelengths}

\author[0000-0002-8310-0554]{Dana E. Anderson}
\affiliation{Department of Astronomy, University of Virginia, 530 McCormick Road, Charlottesville, VA 22904, USA}
\affiliation{Virginia Initiative on Cosmic Origins Fellow}

\author[0000-0003-0787-1610]{Geoffrey A. Blake}
\affiliation{Division of Geological and Planetary Sciences, California Institute of Technology, 1200 E. California Blvd., Pasadena, CA 91125, USA}

\author[0000-0003-2076-8001]{L. Ilsedore Cleeves}
\affiliation{Department of Astronomy, University of Virginia, 530 McCormick Road, Charlottesville, VA 22904, USA}

\author[0000-0003-4179-6394]{Edwin A. Bergin}
\affiliation{Department of Astronomy, University of Michigan, 1085 S. University, Ann Arbor, MI 48109, USA}

\author[0000-0002-0661-7517]{Ke Zhang}
\affiliation{Department of Astronomy, University of Wisconsin-Madison, 475 N. Charter Street, Madison, WI 53706, USA}

\author[0000-0002-6429-9457]{Kamber R. Schwarz}
\affiliation{Lunar and Planetary Laboratory, The University of Arizona, 1629 E. University Blvd., Tucson, AZ 85721, USA}
\affiliation{Sagan Fellow}

\author[0000-0003-3682-6632]{Colette Salyk}
\affiliation{Department of Physics and Astronomy, Vassar College, 124 Raymond Avenue, Poughkeepsie, NY 12604, USA}

\author[0000-0003-4001-3589]{Arthur D. Bosman}
\affiliation{Department of Astronomy, University of Michigan, 1085 S. University, Ann Arbor, MI 48109, USA}

\begin{abstract} \noindent Infrared observations probe the warm gas in the inner regions of planet-forming disks around young sun-like, T~Tauri stars. In these systems, H$_2$O, OH, CO, CO$_2$, C$_2$H$_2$, and HCN have been widely observed. However, the potentially abundant carbon carrier CH$_4$ remains largely unconstrained. The \textit{James Webb Space Telescope} (\textit{JWST}) will be able to characterize mid-infrared fluxes of CH$_4$ along with several other carriers of carbon and oxygen. In anticipation of the \textit{JWST} mission, we model the physical and chemical structure of a T Tauri disk to predict the abundances and mid-infrared fluxes of observable molecules. A range of compositional scenarios are explored involving the destruction of refractory carbon materials and alterations to the total elemental (volatile and refractory) C/O ratio. Photon-driven chemistry in the inner disk surface layers largely destroys the initial carbon and oxygen carriers. This causes models with the same physical structure and C/O ratio to have similar steady state surface compositions, regardless of the initial chemical abundances. Initial disk compositions are better preserved in the shielded inner disk midplane. The degree of similarity between the surface and midplane compositions in the inner disk will depend on the characteristics of vertical mixing at these radii. Our modeled fluxes of observable molecules respond sensitively to changes in the disk gas temperature, inner radius, and the total elemental C/O ratio. As a result, mid-infrared observations of disks will be useful probes of these fundamental disk parameters, including the C/O ratio, which can be compared to values determined for planetary atmospheres. 
\end{abstract}

\keywords{Protoplanetary disks -- Astrochemistry}

\section{Introduction}

(Exo)planetary studies are revealing the chemical composition of planets in our galaxy. Making chemical connections between planets and protoplanetary disks may help us determine the conditions under which these planets formed \citep[e.g.,][]{2016MNRAS.461.3274C}. One promising method of comparison is through measuring elemental abundances, including the ratios of O/H, C/H, and C/O \citep[e.g.,][]{2011ApJ...743L..16O,2019ARA&A..57..617M}. Investigation of the elemental composition of gas in the inner regions of protoplanetary disks, up to a few au from the central star, is of particular interest because this region is thought to be representative of the terrestrial planet forming environment in our own solar system.  

\begin{figure*}
  \includegraphics[width=\linewidth]{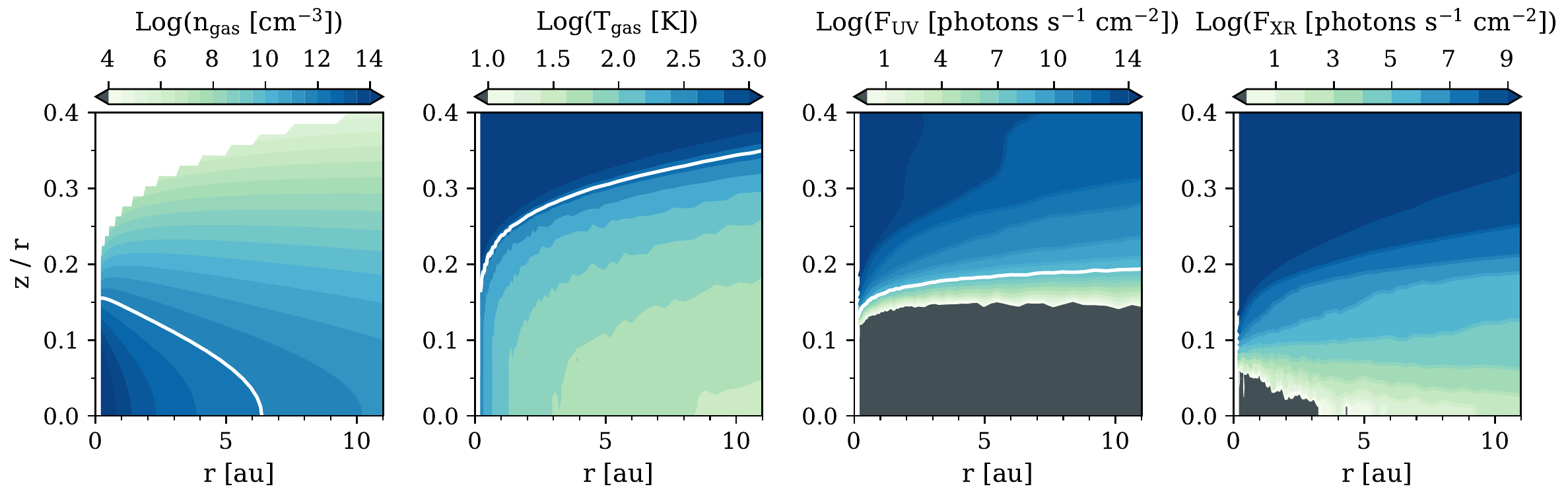}
  \caption{The physical structure of the modeled fiducial disk including the gas density, gas temperature, and UV continuum and X-ray radiation fields. Contours indicate the values of 10$^{12}$~cm$^{-3}$, 500~K, and G$_0$ \citep{1968BAN....19..421H} for reference.}
  \label{fig:env}
\end{figure*}

Gas in the inner disk is mainly observed at near to mid-infrared wavelengths. Various carbon and oxygen carriers are observable in this range. H$_2$O, CO$_2$, HCN, C$_2$H$_2$, and OH have been widely detected in T~Tauri disks by the \textit{Spitzer Space Telescope} \citep[e.g.,][]{2008Sci...319.1504C, 2010ApJ...720..887P, 2011ApJ...731..130S, 2011ApJ...733..102C, 2013ApJ...779..178P}. With Keck-NIRSPEC, the $\nu_3$ band of CH$_4$ at around 3.3 $\mu$m was found in absorption for the disk of T~Tauri star GV~Tau~N \citep{2013ApJ...776L..28G}. The high excitation temperatures derived for these molecules suggest that they are present in the warm inner regions of disks, consistent with the radii associated with terrestrial planet formation. 

In addition to the volatile carriers, a significant source of carbon may be present in solid, refractory form. About half of the total carbon in the interstellar medium exists in a refractory phase \citep[e.g.,][]{1996ARA&A..34..279S, Mishra_2015}. Based on the lack of carbon found in primitive meteorites, roughly 90--99\% of the interstellar refractory carbon is expected to be released into the gas prior to planetesimal formation \citep{2015PNAS..112.8965B}. Various mechanisms have been proposed for the breakdown of refractory carbon in the protostellar and/or protoplanetary disk stages \citep{1997A&A...317..273B, 1997A&A...325.1264F, 2001A&A...378..192G, 2002A&A...390..253G, 2010ApJ...710L..21L,  2017A&A...606A..16G,  2017ApJ...845...13A, 2020ApJ...897L..38V}. Evidence of the release of refractory carbon may be revealed through observations of carbon-bearing species in the disk gas and result in larger C/O ratios \citep{2019ApJ...870..129W}. 

Here we present theoretical models investigating how well emission of observable molecules including H$_2$O, OH, CO$_2$, C$_2$H$_2$, HCN, and CH$_4$ responds to changes in the major carbon and oxygen carriers in the inner regions of protoplanetary disks. The aim of this work is to understand if we can ascertain both the chemical history and bulk composition of the inner disk via near to mid infrared observations, such as those anticipated by the upcoming \textit{James Webb Space Telescope} (\textit{JWST}). We explore several different chemical scenarios to determine how the fluxes of observable species are affected. 

The paper is organized as follows. Section \ref{section2} describes modeling of the protoplanetary disk structure and chemistry. The resulting disk compositions, including a comparison of the observable surface column to the midplane, and the flux predictions for observable chemical species at mid-infrared wavelengths are presented in Section \ref{section3}. The implications of our results are discussed in Section \ref{section4}, followed by a summary of our main conclusions in Section \ref{section5}.

\section{Disk Modeling}\label{section2}
\subsection{Physical Structure}
Our modeled disk surrounds a 1 M$_{\odot}$ T Tauri star with a radius of 2.8 R$_{\odot}$ and an effective temperature of 4300 K. Our fiducial disk model is azimuthally symmetric and has an outer radius of 120 au and a total gas mass of 0.01 M$_{\odot}$. Two populations of dust composed of 80\% astronomical silicates and 20\% graphite \citep{1984ApJ...285...89D} are included. Each population follows a Mathis–Rumpl–Nordsieck (MRN) size distribution \citep{1977ApJ...217..425M}. The large dust population has a maximum size of 1 mm and contains 90\% of the total dust mass. The remaining dust mass exists in the small dust population, with a maximum size of 1~$\mu$m. The minimum dust size for both populations is 0.005~$\mu$m. The dust size distribution is used only in the temperature calculation and the UV and X-ray field calculations. Note that because the opacity coefficient $\kappa$ is defined per unit dust mass, these disk properties are more sensitive to the maximum dust size for each population. The spatial distribution of the disk gas and dust follows that of \citet{1974MNRAS.168..603L}, see also \citet{2011ApJ...732...42A}.  A power law with an exponential cutoff:
\begin{equation}
\Sigma_{gas}(R) = \Sigma_{c}(R/R_{c})^{-1}exp[-(R/R_{c})] 
\end{equation}
where $\Sigma_{c}$ =  23.0~g~cm$^{-2}$ and $R_{c}$ = 30~au, describes the fixed radial surface density distribution of gas. For simplicity, the dust surface densities are assumed to have the same structure with 100$\times$ lower $\Sigma_{c}$. The surface density is chosen to normalize the total disk mass to 0.01~M$_{\odot}$, which is close to the minimum mass solar nebula. The scale height of both the gas and the small dust population is 10 au at a radius of 100 au and is distributed radially with a power-law of $H/R$ $\propto$ $R^{0.15}$. The scale height of the large dust population is 5$\times$ lower in order to approximate dust settling. This results in a gas-to-dust ratio of 1000 in the surface layers where only small dust resides while the vertically integrated gas-to-dust ratio is 100. 

The disk is passively heated by the central star and dust temperatures are calculated using TORUS \citep{2004MNRAS.350..565H}. The Monte Carlo radiative transfer code of \citet{2011ApJ...740....7B,2011ApJ...739...78B} is used to compute the UV and X-ray radiation fields using the UV spectrum from TW Hya and a total X-ray luminosity of 10$^{29.5}$~erg~s$^{-1}$ between 1--20~keV  as inputs. Gas temperatures are estimated based on (1) the local UV flux and (2) the gas density using fitting functions to the thermochemical models of \cite{2013A&A...559A..46B}, as described in \cite{2015ApJ...799..204C}. For cosmic ray ionization, we used the Solar System Minimum model from \cite{2015ApJ...799..204C}. The incident rate at the disk surface is 1.1$\times$10$^{-18}$ s$^{-1}$ and is modulated with vertical depth in the disk. Figure \ref{fig:env} shows the structure of the modeled disk. 

\begin{table}[b]
\caption{Initial Abundances}
\begin{tabular}{llll}
\hline 
H$_2$ &  5.000(-01) & HCN   &  2.000(-08) \\
He  &  1.400(-01) & C$_2$H   &  8.000(-09)\\
H$_2$O  &   8.000(-05) &  SO   &  5.000(-09)\\
CO   &  9.920(-05) &  CS  &  4.000(-09)\\
CO$_2$  &  2.240(-05) &  H$_3^+$  & 1.000(-08)\\
CH$_3$OH & 4.800(-06) &  HCO$^+$  &  9.000(-09)\\
CH$_4$   &  3.600(-06) & Si$^+$ & 1.000(-11)\\
N$_2$  &  3.510(-05) &  Mg$^+$ & 1.000(-11)\\
NH$_3$  &  4.800(-06) &  Fe$^+$  &  1.000(-11) \\
\hline
\multicolumn{4}{l}{Note: a(b) = a$\times$10$^{\mathrm{b}}$ per total H}
\end{tabular}
\label{table:init}
\end{table}

\subsection{Chemical Model}
The physical conditions present in cold, interstellar environments limits viable chemistry to exothermic reactions with no activation energy barriers. However, certain reactions that are slightly endothermic or have moderate activation energy barriers, so-called high-temperature reactions, may become important in protoplanetary disks where temperatures can reach 100s to 1000s of K near the central star and/or at the disk surface. Infrared observations probe these warmer disk regions; therefore, including high-temperature reactions in our network is crucial. We incorporated high-temperature gas-phase reactions (appropriate for up to 800 K) from \cite{2010ApJ...721.1570H} into the gas-grain chemical reaction network of \cite{2018ApJ...865..155C}, used for the interpretation of sub-mm/mm observations of molecular species in protoplanetary disks. We also updated the model's photodissociation cross section data based on \cite{2017A&A...602A.105H}. More details regarding these modifications to the chemical model are provided in the Appendix.
 
 The initial abundances assumed for the fiducial model are given in Table~\ref{table:init}. Major volatile abundances are based on interstellar ice compositions \citep[][]{2015ARA&A..53..541B}. Because the model is dynamically static, the outcomes of mass transport within the disk are approximated by varying these initial abundances as described in Section \ref{section3}. In models where refractory carbon is included, destruction via oxidation and UV photolysis are added to the reaction network using the method described for carbonaceous grains by \cite{2017ApJ...845...13A}. Two types of refractory carbon reactions are added to the network: (1) Oxidation of carbon atoms on refractory grain surfaces by O producing CO, with a reaction rate from \citet{2010ApJ...710L..21L} and (2) UV photolysis of hydrogenated amorphous carbon surfaces releasing CH$_4$ in addition to minor amounts of C$_2$H$_2$, C$_2$H$_4$, and C$_2$H$_6$ based on the laboratory measurements of \citet{2014A&A...569A.119A, 2015A&A...584A.123A}. The carbonaceous grains are assumed to be spherical with a radius of 0.1 $\mu$m and composed of pure carbon in the case of oxidation or hydrogenated amorphous carbon in the case of UV photolysis. The amount of carbon initially stored in refractory form is 3.33$\times$10$^{-5}$ relative to total H, equivalent to roughly 40\% of the mass of the small dust population. The chemical models are run for a timescale of 3$\times$10$^6$ years.

\subsection{Observational Predictions}\label{section2.3}
We compute fluxes for CO$_2$, C$_2$H$_2$, HCN, H$_2$O, OH, and CH$_4$ using the Python code \texttt{slabspec}, the local thermal equilibrium (LTE) emission model of \cite{2011ApJ...731..130S}. The code is available on GitHub\footnote{Supplementary materials:~\url{http://github.com/csalyk/slabspec}.} and archived in Zenodo \citep{slabspec_key}. \texttt{slabspec} receives its molecular line parameters from the HITRAN database \citep{2017JQSRT.203....3G}. The model uses the slab approximation and takes three inputs: column density, gas temperature, and area. A Gaussian line shape where $\sigma$~=~2~km~s$^{-1}$ is assumed, the same value used by \cite{2011ApJ...731..130S} for fitting Spitzer and Keck-NIRSPEC data. 

The slab model is run for a series of annuli corresponding to $\sim$70 radial grid points in our disk model from 0.2 to 10~au. For each radial slice, the vertical column of a given molecular species is integrated from the disk surface down to a total H$_2$ gas column density of 1.8$\times$10$^{23}$~cm$^{-2}$.
\begin{figure*}[t]
    \includegraphics[scale=0.75, trim={0.2cm 0cm 0cm 0cm}, clip]{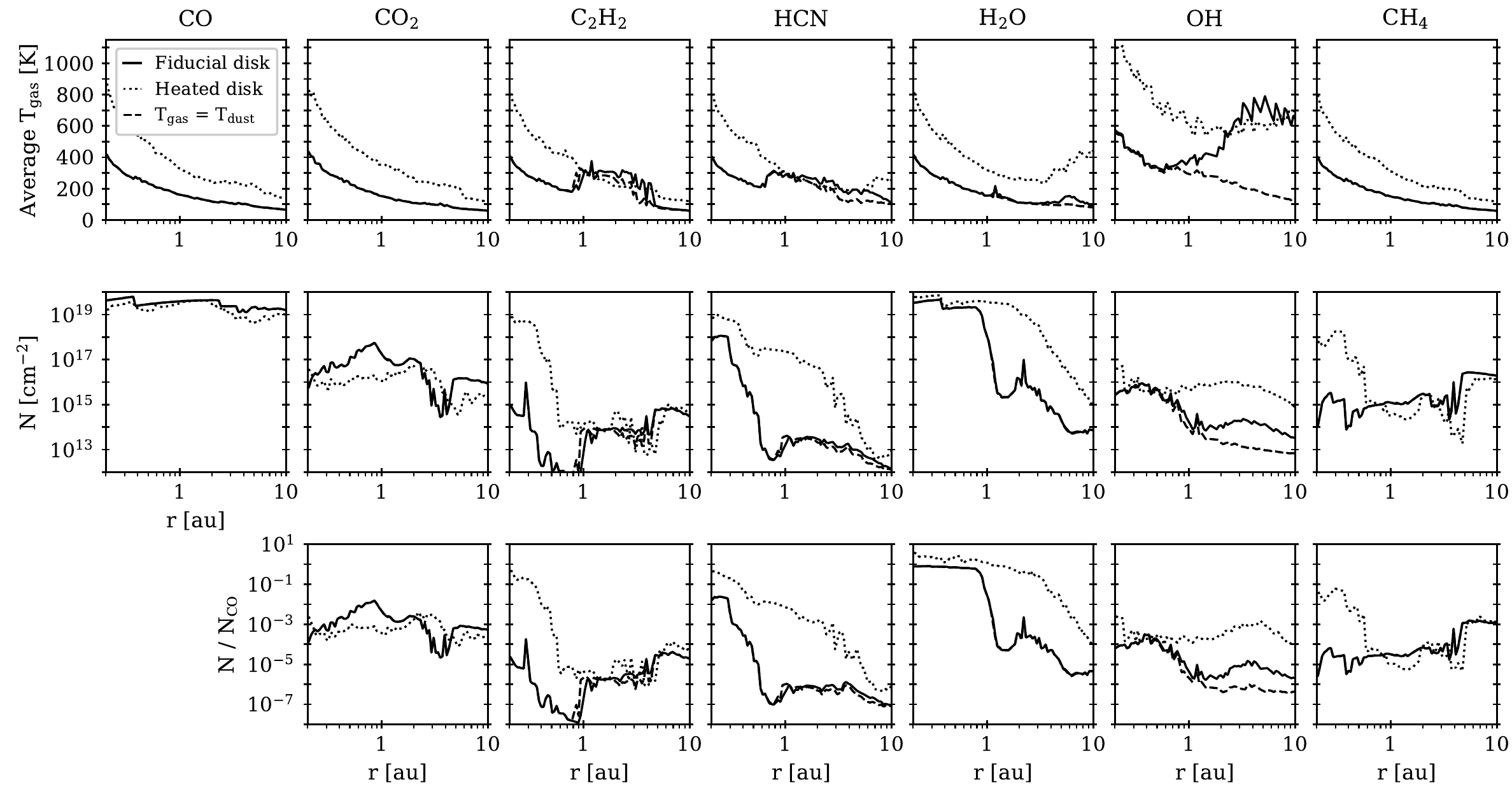}
    \caption{{Surface density weighted average gas temperature (top row), surface column density (middle row), and surface column density relative to that of CO (bottom row) versus distance from the central star for observable molecules at 10$^6$ years in the fiducial model (indicated with the solid line), a model where disk gas temperatures were increased by a factor of 2 (dotted line), and a model where the gas and dust are thermally coupled (dashed line).}} 
\label{fig:NvR}
\end{figure*}
This threshold is selected as the location where the dust becomes optically thick at infrared wavelengths for a gas-to-dust ratio of 1000 in the disk surface layers given the adopted opacities for our dust mixture. We assume a face-on disk orientation, excluding the effects of disk inclination. An average gas temperature weighted by the surface density profile for each molecular species in the vertical (z) direction is used as the input temperature for each annulus. 

The flux densities for all of the individually modeled annuli are summed to produce a final spectrum. Single value reported fluxes are integrated over the wavelength ranges used by \citet{2011ApJ...731..130S} for \textit{Spitzer} observations of disks (Table~\ref{table:wv}). The CH$_4$ flux density is integrated over a more generous range, from 5.0 to 11.0~$\mu$m. We ignore absorption of midplane continuum emission by surface molecules, which could cause the observed fluxes of optically thick lines to be fainter than those reported here.

\section{Results}\label{section3}

 \subsection{Molecular Column Densities and Temperatures}
Figure \ref{fig:NvR} shows the surface column densities and the surface density weighted average gas temperatures versus disk radius for several molecules that are of interest as potential targets for \textit{JWST} Mid-Infrared Imager (MIRI) Medium Resolution Spectroscopy \citep[MRS;][]{2015PASP..127..646W} observations. Fig.~\ref{fig:NvR} compares the fiducial disk model described in Section \ref{section2} to a model where the gas temperature is set equal to that of the dust and a ``heated" disk model where the gas temperature is increased by a factor of two throughout the disk prior to the chemical calculations. The additional models are included to estimate the effect of the gas temperature, including a much warmer gas temperature, on the observable emission. The higher temperatures are more in line with those derived from infrared observations of several molecules by \cite{2011ApJ...733..102C} and \cite{2011ApJ...731..130S}.

\begin{table}[b]
\caption{Selected IR Spectral Regions}
\centering
\begin{tabular}{lc}
\hline
& Wavelength [$\mu$m] \\
\hline 
CO$_2$ &          14.847--15.014\\
C$_2$H$_2$     &     13.553--13.764\\
HCN   &      13.837--14.075\\
H$_2$O    &      17.190--17.260\\
OH    &      23.009--23.308 \\
& 27.308--27.764\\
CH$_4$     &     5.000--11.000\\
\hline
\end{tabular}
  \label{table:wv}
\end{table}

When considering the gas column integrated vertically into the disk to a total H$_2$ column density of 1.8$\times$10$^{23}$~cm$^{-2}$ (here the ``surface column"), the average gas temperature at which most of these molecular species reside decreases with radius from about 400 K at 0.2 au to 100 K at 5 au for the fiducial model or 800--900 and 200--300 K, respectively, for the heated disk (Fig.~\ref{fig:NvR}, top row). At larger radii the average gas temperatures for HCN and H$_2$O begin to increase. Relative to the other molecular species shown here, OH abundances are concentrated in higher temperature regions closer to the disk surface and are the most affected when gas temperatures are not assumed to be enhanced relative to the dust. 

Peak surface column densities range from around 10$^{15}$--10$^{19}$ cm$^{-2}$ and vary among different molecular species. In most cases, the surface column density is higher closer to the central star and drops off at larger radii. For HCN, C$_2$H$_2$, H$_2$O, CO$_2$, and CH$_4$, this behavior is common among multiple models of inner disk chemistry \citep[e.g.,][]{2011ApJ...743..147N,2015A&A...582A..88W,2018A&A...616A..19A}. In the fiducial model, the surface column densities of C$_2$H$_2$ and CH$_4$ do not peak significantly at smaller radii and therefore differ from the radial behavior that is typically seen. 

The increased gas temperatures in the heated disk model result in higher column densities for H$_2$O and OH, mostly at radii beyond 1 au as H$_2$O freezeout is prevented. C$_2$H$_2$ and CH$_4$ column densities also increase, but interior to 1~au, and HCN column densities are higher at all radii within 10~au for the heated disk. \citet{2008A&A...483..831A} describe the chemistry behind the enhanced production of small organic species at high temperatures. Bimolecular reactions of small carbon-bearing species (e.g., C$^+$, C, CN, and C$_2$H) with H$_2$, which lead to the formation of CH$_4$, HCN, and C$_2$H$_2$, have high activation energy barriers. High gas temperatures, above 400--700 K, are required to activate these reactions. 

\cite{2011ApJ...731..130S} derived excitation temperatures $>$450--500~K and up to 1200--1600~K from infrared observations of CO, CO$_2$, C$_2$H$_2$, HCN, H$_2$O, and OH in 48 T~Tauri disks. Typical observed CO and H$_2$O column densities were 10$^{18}$--10$^{19}$~cm$^{-2}$ while those of CO$_2$, C$_2$H$_2$, HCN, and OH were in the range of 10$^{14}$--10$^{16}$~cm$^{-2}$ with a ratio of about 10$^{-3}$ relative to H$_2$O and CO. Using a slightly different fitting approach, \cite{2011ApJ...733..102C} found line-of-sight column densities in roughly this same range for H$_2$O, HCN, and C$_2$H$_2$ with derived temperatures of 500--800~K based on \textit{Spitzer} observations of 11 T~Tauri disks. Their best-fit temperatures for CO$_2$ were lower: 100--600~K with an average of 350~K, suggesting that CO$_2$ emission originates in a cooler region radially or vertically in the disk. We do not expect to match the best-fit column densities and temperature values from these works exactly because of the differences in our modeling approaches. \cite{2011ApJ...731..130S} and \cite{2011ApJ...733..102C} identify a single column density and temperature over a specified area in the inner disk, where \cite{2011ApJ...731..130S} uses the estimated emitting area of H$_2$O and \cite{2011ApJ...733..102C} fit this area separately for different molecules. In comparison, this work models a series of annuli with individual column density and temperature values for disk radii from 0.2--10 au.  Nevertheless, it is worth noting that our results generally agree with those derived from observations in that the gas is rich in CO and H$_2$O with somewhat lower column densities of CO$_2$, C$_2$H$_2$, HCN, and OH. 

\cite{2013ApJ...776L..28G} observed CH$_4$ in absorption for the disk around GV~Tau~N at near-IR wavelengths with a column density of (2.8$\pm$0.2)$\times$10$^{17}$ cm$^{-2}$. The near-IR column density was 2$\times$ higher than that of HCN in the same source, with derived rotational temperatures of 750 K for CH$_4$ and 550 K for HCN. The difference in rotational temperatures suggests that CH$_4$ abundances peak closer to the central star. As shown in Fig.~\ref{fig:NvR}, the level of CH$_4$ in our fiducial model is $\sim$10$^{15}$ cm$^{-2}$. Higher disk gas temperatures are needed to reach the observed value. Although the peak HCN column density is higher than that of CH$_4$, the heated disk model does indicate that CH$_4$ is concentrated at smaller radii whereas large column densities of HCN are present out to several au from the star in agreement with the general behavior noted for GV~Tau~N.  

\begin{figure*}[t]
  \includegraphics[width=\linewidth]{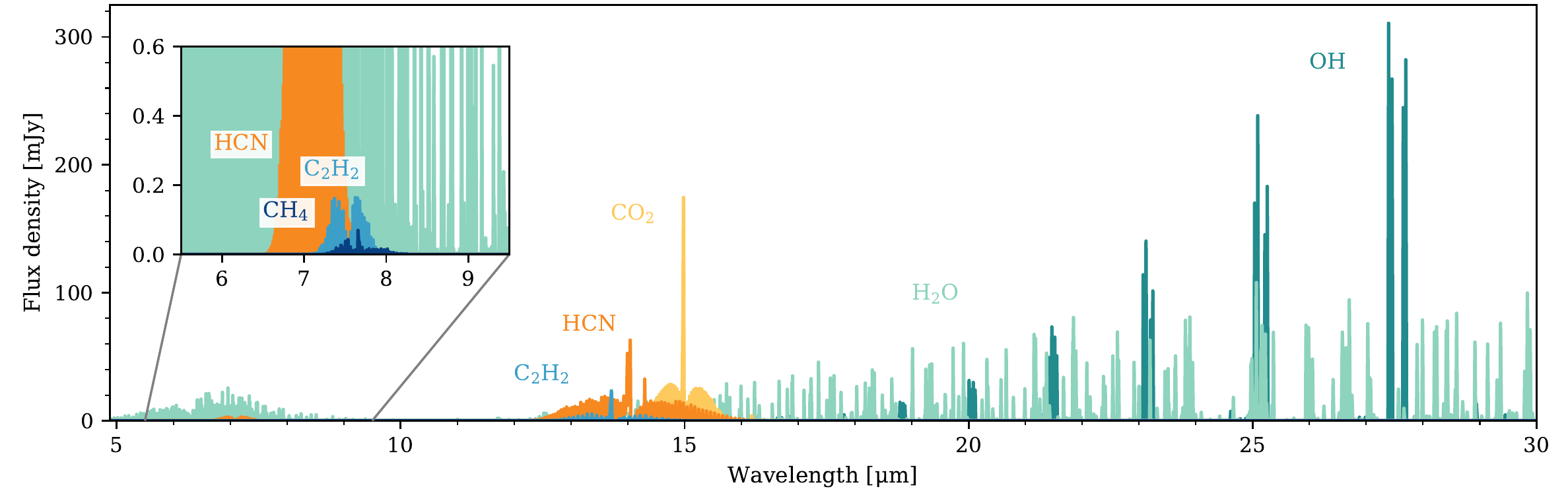}
  \caption{Modeled spectra for CH$_4$, C$_2$H$_2$, HCN, CO$_2$, H$_2$O, and OH from the fiducial disk model at 10$^6$ years and a distance of 140 pc. Spectra are convolved to a spectral resolving power of R=3000, similar to the resolving power of \textit{JWST}-MIRI MRS. CH$_4$ emission is shown in the inset.}
  \label{fig:spectrum}
\end{figure*}

\begin{figure}
  \includegraphics[width=\linewidth]{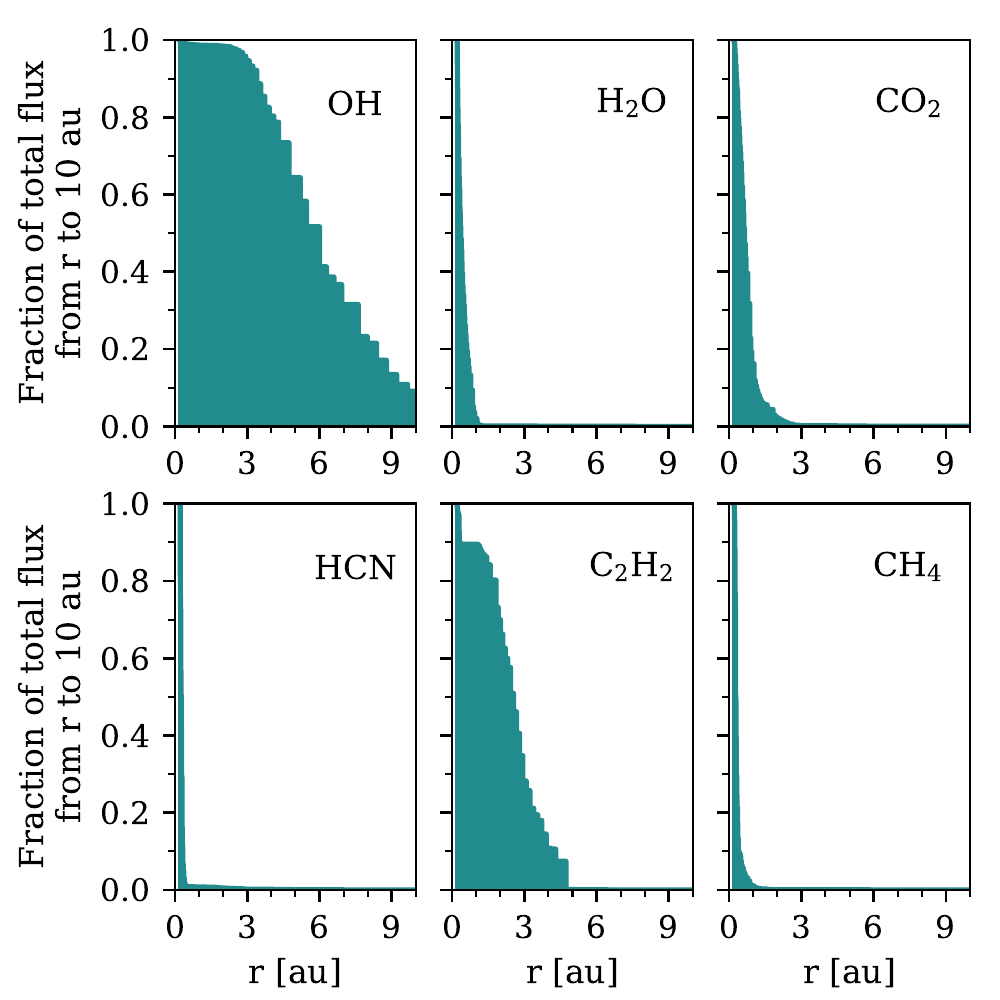}
  \caption{Radial origin of disk emission for molecular species from the fiducial disk model at 10$^6$ years. The fluxes modeled for individual annuli are summed over all wavelengths from 4.88 to 30~\micron~and for radii between r and 10~au then plotted relative to the total flux for all annuli (0.2--10~au).}
  \label{fig:rad}
\end{figure}

\begin{figure}
    \includegraphics[width=\linewidth]{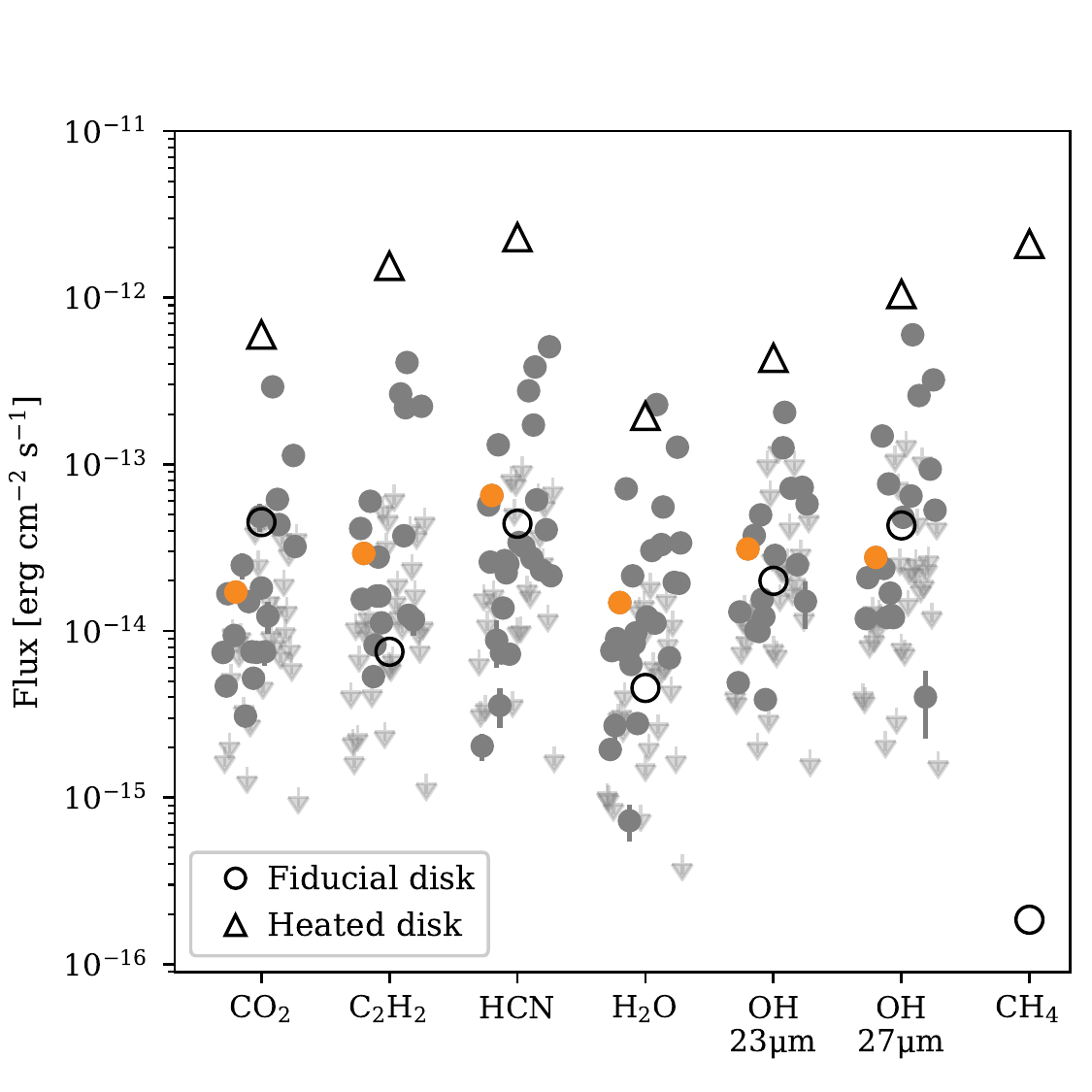}
    \caption{{Modeled fluxes for the fiducial model at 10$^6$ years and 140~pc integrated over the wavelength ranges in Table~\ref{table:wv} (indicated with open circles). The fiducial model fluxes are compared to those from a model where the gas temperatures were increased by a factor of 2 (triangles). Plotted in gray are the observed fluxes from a survey of T Tauri disks by \cite{2011ApJ...731..130S} for the same molecules and wavelength ranges scaled to a source distance of 140~pc. Downward arrows represent upper limits. Fluxes for AA Tau are shown in orange to indicate an example of the variation within a single source.}} 
\label{fig:flux_obs}
\end{figure}
\begin{figure}[t]
    \includegraphics[width=\linewidth, trim={0 0 0 10mm}, clip]{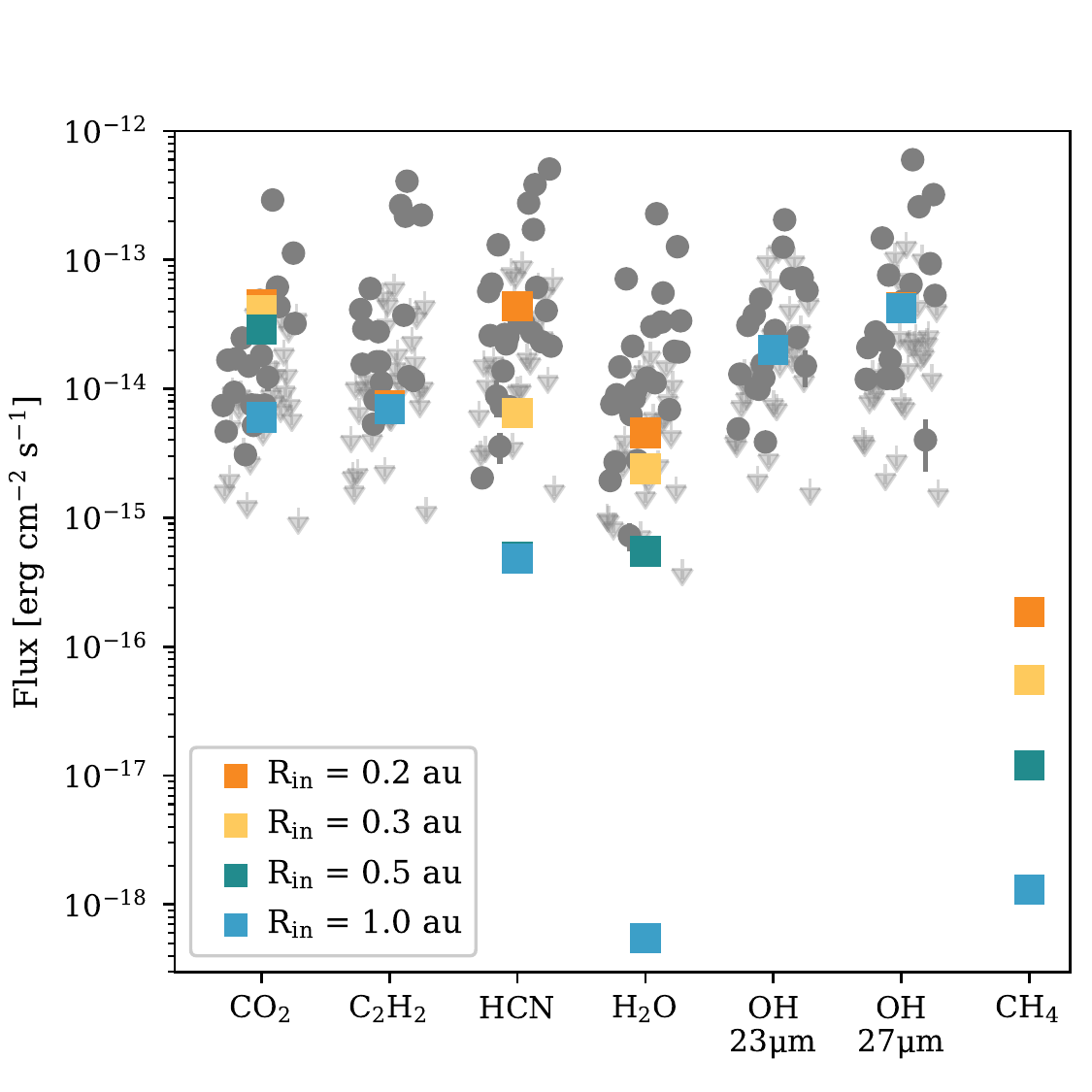}
    \caption{{Comparison of fluxes for different inner gas radii in the fiducial model at 10$^6$~years and 140~pc integrated over the wavelength ranges in Table~\ref{table:wv}. Colors indicate different values for the radius of the inner disk gas edge from 0.2--1.0~au. Fluxes were computed by summing the modeled flux per radial bin from the inner edge out to 10~au. Plotted in gray are the observed fluxes from a survey of T Tauri disks by \cite{2011ApJ...731..130S} for the same molecules and wavelength ranges scaled to a source distance of 140~pc. Downward arrows represent upper limits.}} 
\label{fig:flux_rin}
\end{figure}

\vspace{1cm}
\subsection{Predicted Spectra}
The modeled spectra for CH$_4$, C$_2$H$_2$, HCN, CO$_2$, H$_2$O, and OH from 4.88--30.00 $\mu$m, a spectral resolving power R=3000, and the fiducial disk model are provided in Figure~\ref{fig:spectrum}. For the fiducial model, the HCN, H$_2$O, and CH$_4$ emission originates within 1~au from the central star. Some of the CO$_2$ and C$_2$H$_2$ emission comes from larger radii, out to 3--5~au. In contrast, a substantial portion of the OH emission comes from radii beyond a few au from the central star (Figure~\ref{fig:rad}). For comparison, \cite{2011ApJ...733..102C} derived emitting areas with a radius of about 1~au for H$_2$O, 0.2--0.6~au for HCN, and $>$0.5--2.0~au for CO$_2$.

Figure~\ref{fig:flux_obs} shows the modeled fluxes for C$_2$H$_2$, HCN, CO$_2$, H$_2$O, OH, and CH$_4$ integrated over the spectral ranges given in Table~\ref{table:wv}. The outcomes from the fiducial and heated disk models are compared. Increasing fiducial disk gas temperatures by a factor of 2 in the heated disk model causes fluxes to increase by at least an order of magnitude for all six of the molecular species. The CH$_4$ flux experiences the most extreme difference due to the enhanced temperature. Relative to the molecules that have been detected by \textit{Spitzer}, the predicted flux for CH$_4$ is similar for the heated disk model but significantly lower for the fiducial disk. 

\begin{figure}[b]
    \includegraphics[width=\linewidth]{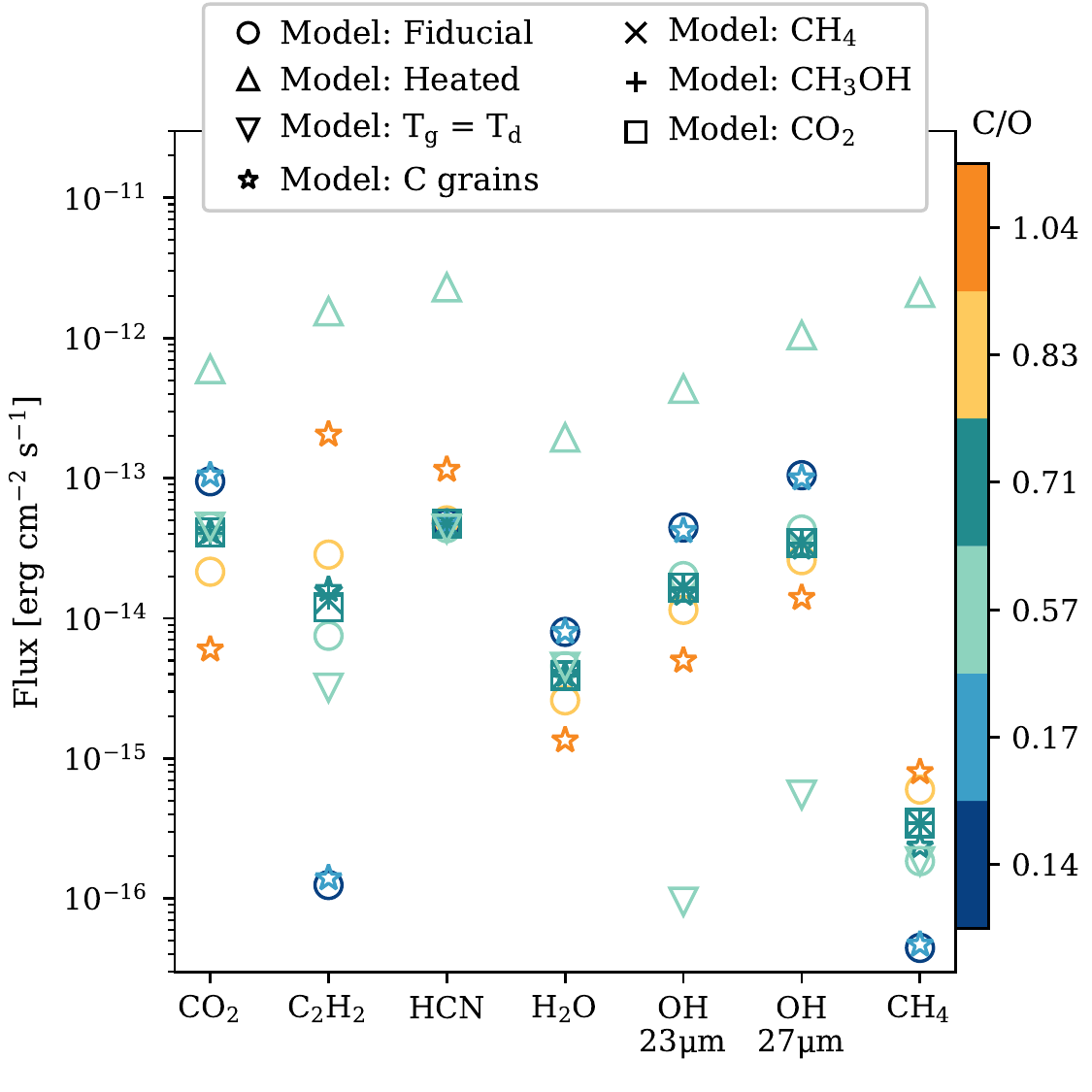}
    \caption{{Comparison of fluxes for several models at 10$^6$~years and 140~pc integrated over the wavelength ranges in Table~\ref{table:wv}. Colors indicate C/O ratios and symbols indicate the following initial conditions: models based on the fiducial setup where the C/O ratio is determined by varying the H$_2$O abundance (indicated with open circles, note that the composition in Table~\ref{table:init} corresponds to a C/O ratio of 0.57), models including the destruction of refractory carbon grains (stars), three additional models with a C/O ratio of 0.71 but different relative amounts of major carbon carriers (`x', `+', and square, see Section~\ref{section3.3} of the text for more details), and models where the disk gas temperature was increased by a factor of 2 (upward triangles) or set equal to the dust temperature (downward triangles) and the initial composition is given in Table~\ref{table:init}.}} 
\label{fig:flux_C2O}
\end{figure}

In comparison to the sample of T Tauri disk fluxes observed by \cite{2011ApJ...731..130S}, the fiducial model fluxes appear within the range of detected values but are relatively high compared to many of the upper limits (Fig.~\ref{fig:flux_obs}). This is also true in comparison to the fluxes found by \cite{2011ApJ...733..102C} although the wavelength ranges used to calculate the integrated fluxes differ slightly. Meanwhile, the heated disk fluxes exceed the observed ranges. For HCN, CO$_2$, and H$_2$O, which have lines that become optically thick at select disk radii in the fiducial model (and all molecules in the heated disk model), absorption of continuum photons at wavelengths where the lines are optically thick may cause observed fluxes to be smaller than shown here after a standard continuum subtraction is performed. Additionally, line fluxes are affected by the assumed inner gas radius of the disk. Increasing this radius from 0.2~au to 1.0~au significantly lowers the fluxes of some molecules whereas the fluxes of C$_2$H$_2$ and OH, both of which have substantial contributions of flux from larger radii (Fig.~\ref{fig:rad}), remain largely unaffected (Figure~\ref{fig:flux_rin}). 

\subsection{Varying Disk Compositions}\label{section3.3}
We now test the dependence of the predicted fluxes on the initial chemical composition of the disk. First, we changed the elemental C/O ratio in the model by altering the initial H$_2$O abundance. Relative to the disk composition provided in Table~\ref{table:init} with an elemental C/O ratio for volatile materials of 0.57, the initial H$_2$O abundance was increased by a factor of 10 resulting in a C/O ratio of 0.14 and decreased by a factor of 10 for a C/O ratio of 0.83. With the exception of HCN, the predicted fluxes are sensitive to these changes in the C/O ratio with C$_2$H$_2$ displaying the widest range of fluxes (Figure~\ref{fig:flux_C2O}). The difference in sensitivity between C$_2$H$_2$ and HCN reflects a difference in modeled optical thickness for the wavelength regions investigated (given in Table~\ref{table:wv}). 

Next, we changed the elemental C/O ratio by including the destruction of a source of refractory carbon with an initial abundance of 3.33$\times$10$^{-5}$ C atoms relative to total H. This results in a change in the C/O ratio from 0.57 in the fiducial model to 0.71 in the model with carbon grain destruction. Refractory carbon is removed from grain surfaces either by reacting with an oxygen atom at temperatures above a few hundred~K producing CO or by a UV photon releasing CH$_4$ \citep[assuming a starting material of hydrogenated amorphous carbon;][]{2015A&A...584A.123A}. The surface regions interior to 10~au studied here represent an area of the disk where 0.1~$\mu$m carbon grains are completely destroyed within the disk lifetime. Therefore, by 10$^6$ years, the elemental C/O ratio of volatile materials is equal to the total (refractory + volatile) elemental C/O ratio for the model. 

As Fig.~\ref{fig:flux_C2O} shows, the inclusion of carbon grains in the model generally causes the fluxes of C$_2$H$_2$, HCN, and CH$_4$ to increase and the fluxes of oxygen-bearing species to decrease. This pattern matches that found when increasing the C/O ratio by decreasing the initial H$_2$O abundance. The effect of including carbon grains on the predicted fluxes increases for lower initial H$_2$O abundances, making it easier to distinguish the models with and without carbon grains at high C/O ratios. 

Changing the disk gas temperatures by a factor of 2 produces larger flux values than altering the C/O ratio within the range of 0.14--1.04 for all molecular species. This is the result of both high-temperature chemical changes as seen in the heated disk surface column densities in Fig.~\ref{fig:NvR} and higher excitation temperatures. The range in predicted fluxes for the molecules shown here is about one order of magnitude for the heated disk, lower than that for the other modeled scenarios where differences among the fluxes for different molecular species are more pronounced. Assuming that the gas temperatures are equal to the dust temperatures decreases the modeled fluxes for C$_2$H$_2$ and OH while other molecular fluxes are not affected.

Finally, we modeled four different initial disk compositions with the same elemental C/O ratio of 0.71. Compared to the fiducial disk composition (Table~\ref{table:init}), an excess of 3.33$\times$10$^{-5}$ C atoms relative to total H was added in the form of (1) refractory carbon grains as described above, (2) CH$_4$, (3) CH$_3$OH, and (4) CO$_2$. In the cases of CH$_3$OH and CO$_2$, the initial H$_2$O abundance was reduced relative to the fiducial model in order to achieve a C/O ratio of 0.71. The predicted fluxes are largely the same among these four models (Fig.~\ref{fig:flux_C2O}) indicating a lack of sensitivity to the relative abundances of initial carbon carriers.

\begin{figure}[t]
    \includegraphics[width=\linewidth]{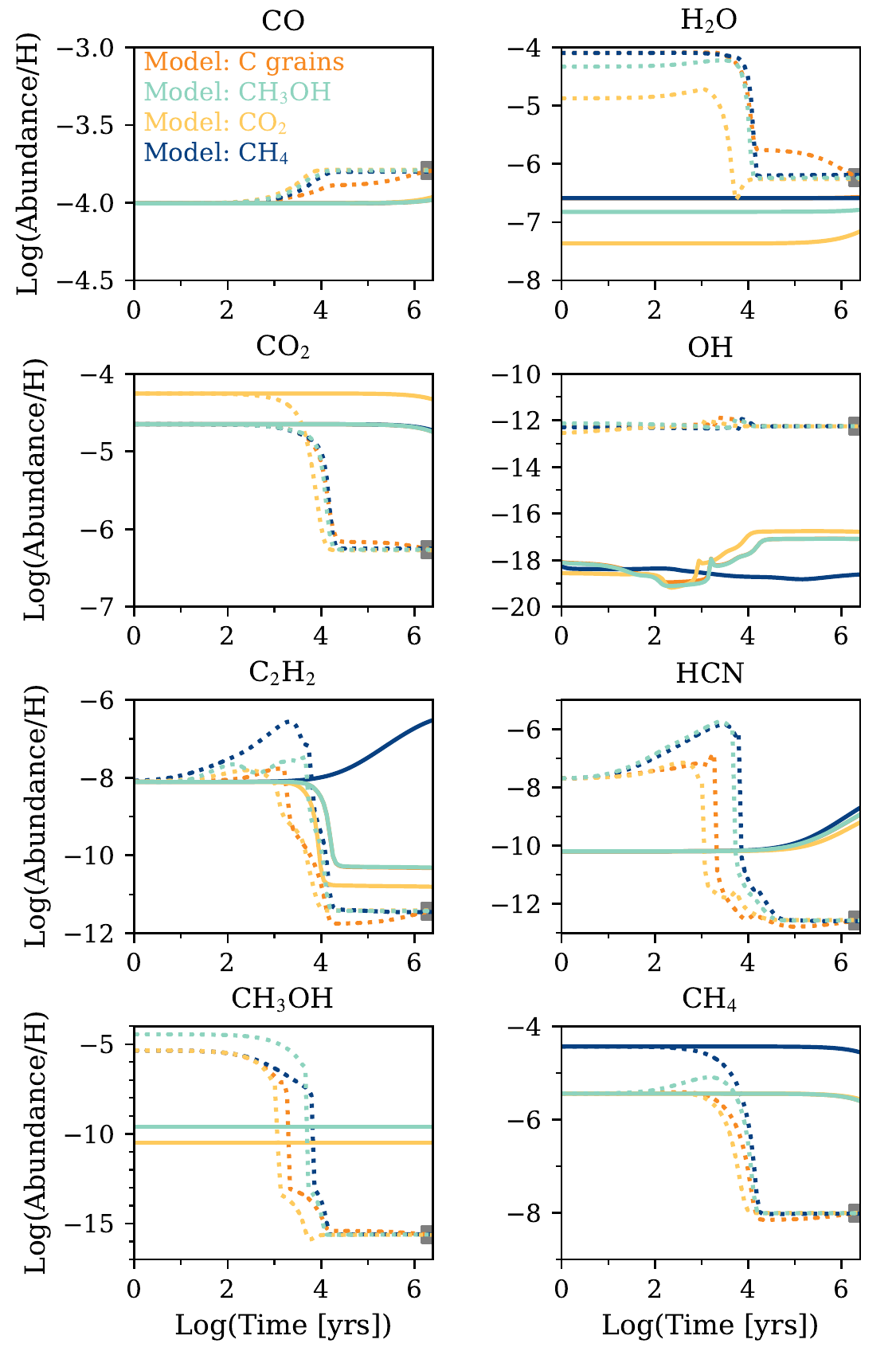}
    \caption{{The fractional abundance of several chemical species relative to total H over time plotted for four modeled scenarios with the same total elemental C/O ratio but different initial compositions (see Section~\ref{section3.3}). Abundances are compared for two disk locations at 1~au from the central star: the midplane (solid lines) and 0.15 au above the midplane (dotted lines). Gray squares highlight the steady-state surface abundance to which all of the models converge.}} 
\label{fig:mid}
\end{figure}

\subsection{Surface vs.~Midplane Composition}

Here we compare the composition of the disk surface to that of the midplane over time. Figure~\ref{fig:mid} shows this comparison at 1~au from the central star for the models exploring different relative amounts of initial carbon carriers (Section~\ref{section3.3}). At the final simulation timestep of  3$\times$10$^6$ years, the chemical composition of the disk surface gas converges for all four models regardless of the initial distribution of elemental abundances among molecular species. When excess carbon is introduced in the form of CH$_4$, CH$_3$OH, or CO$_2$ in photon-dominated regions near the disk surface, these molecules are quickly destroyed by photodissociation or reactions with ionized species. Here ionized species are largely generated as a result of the ionization of H$_2$ and He via X-rays. Carbon is eventually transferred to CO. Meanwhile, oxygen is transferred from H$_2$O, CO$_2$, and CH$_3$OH to CO, O, and O$_2$. Note that the relative importance of H$_2$O gas and ice, CO, O, and O$_2$ as oxygen carriers in this surface layer depends on the disk radius. 

In models that include destruction of carbon grains, UV photolysis at the disk surface releases CH$_4$, which is then converted to CO. In  atomic O-rich regions where the temperature is above a few hundred~K, oxidation of refractory carbon can also produce CO directly. For the location shown in  Fig.~\ref{fig:mid}, carbon release occurs more slowly from the refractory form than from CH$_4$, CH$_3$OH, or CO$_2$ causing the carbon grain model to take somewhat longer to equilibrate. 

In the case of the disk surface layers at 1~au: regardless of the initial carbon carriers, more than 99\% of the available carbon ultimately ends up being stored in CO. For most of the molecular species shown in Fig.~\ref{fig:mid}, steady state abundances are reached around $\sim$10$^4$ years. The specific conversion timescale will depend on the location within the disk and the local UV radiation environment. 

Midplane abundances of the major carbon carriers (CO, CO$_2$, CH$_4$, and CH$_3$OH) experience relatively less and in some cases no change over a timescale of 10$^6$ years at 1 au (Fig.~\ref{fig:mid}, solid lines). Unlike in the surface layers, final abundances of the major carbon carriers in the midplane tend to reflect the initial composition provided to the model. The reason for this slower chemistry is that in denser regions, shielded from UV and X-ray photons, a combination of cosmic-ray ionization and ion chemistry together control the destruction of CH$_4$, CH$_3$OH, and CO$_2$. Therefore, midplane chemistry is largely dependent on the degree of ionization in these dense disk regions. The cosmic-ray ionization rate is not well constrained for most protoplanetary disks. The interstellar value for ionization of H$_2$ is around 3$\times$10$^{-17}$~s$^{-1}$, but may be lower due to the stellar winds or magnetic fields of young, accreting T Tauri stars. The H$_2$ cosmic-ray ionization rate has been estimated to be $\lesssim$10$^{-19}$~s$^{-1}$ based on observations of ionized species for the disk of TW Hya \citep{2015ApJ...799..204C}. 

As a test case, we started our model with all of the initial volatile carbon in the form of CH$_3$OH ice. At 10$^6$ years and 1~au in the disk midplane, around 20\% of carbon was transferred to other chemical species when using an interstellar cosmic-ray ionization rate whereas none of the carbon was removed from the original carrier for an H$_2$ cosmic-ray ionization rate of 10$^{-19}$~s$^{-1}$. These findings of a halted midplane chemistry at low cosmic-ray ionization rates are consistent with those of \citet{2014ApJ...794..123C, 2018A&A...613A..14E} and \citet{2020ApJ...890..154P}.

\begin{figure}[t]
    \includegraphics[width=\linewidth]{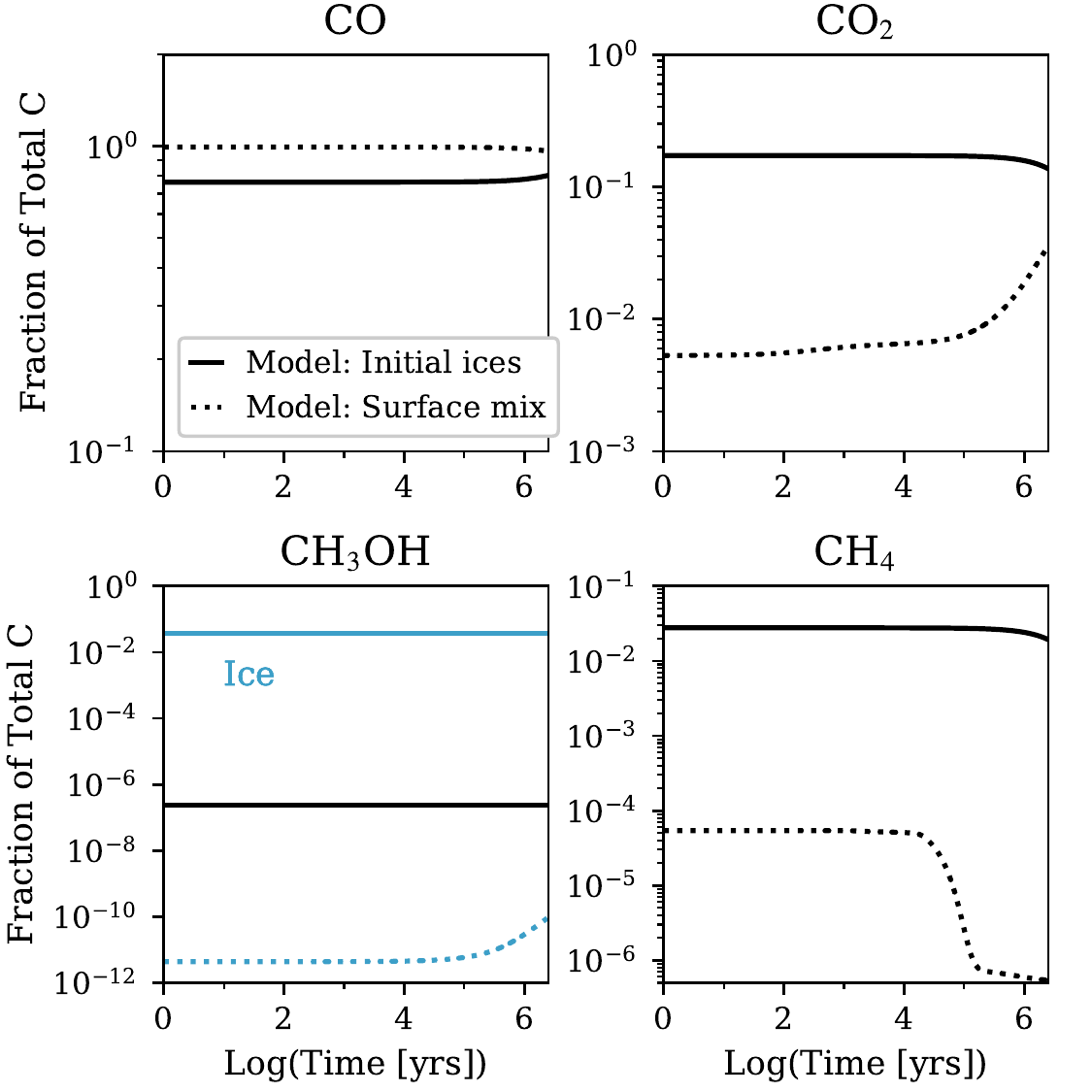}
    \caption{{Fraction of the total carbon over time in CO, CO$_2$, CH$_3$OH, and CH$_4$ in the midplane at 1~au. Models with different initial compositions are compared including the fiducial disk composition (Table~\ref{table:init}, solid line) and a disk surface-like composition (Table~\ref{table:surf}, dotted line). Ice abundances are shown in blue.}} 
\label{fig:surf2mid}
\end{figure}

Finally, we explore what would happen if due to vigorous mixing, the chemical composition of the surface is introduced into the midplane. Essentially, we want to see if the surface composition will return to a state that looks like that of a static midplane model. Figure~\ref{fig:surf2mid} compares midplane chemical abundances for the static model initiated with interstellar ices (Table~\ref{table:init}), to a model where the initial composition reflects the disk surface composition at a height of 0.15~au above the disk midplane for a radius of 1~au (Table~\ref{table:surf}). 

\begin{table}[]
\caption{Disk Surface Abundances at 1~au$^a$}
\begin{tabular}{llll}
\hline 
H$_2$ &  5.000(-01) & HCN   &  2.000(-08) \\
He  &  1.400(-01) & C$_2$H   &  8.000(-09)\\
H$_2$O  &   1.753(-06) &  SO   &  5.000(-09)\\
CO   &  1.293(-04) &  CS  &  4.000(-09)\\
CO$_2$  &  6.854(-07) &  H$_3^+$  & 1.000(-08)\\
CH$_3$OH & 4.000(-16) &  HCO$^+$  &  9.000(-09)\\
CH$_4$   &  7.026(-09) & Si$^+$ & 1.000(-11)\\
N$_2$  &  3.738(-05) &  Mg$^+$ & 1.000(-11)\\
NH$_3$  &  1.586(-09) &  Fe$^+$  &  1.000(-11) \\
O$_2$  &  4.015(-05) &   O  &  1.606(-05) \\
N  &  2.601(-07)  \\
\hline
\multicolumn{4}{l}{Note: a(b) = a$\times$10$^{\mathrm{b}}$ per total H} \\
\multicolumn{4}{l}{$^a$For the fiducial model at z = 0.15~au} 
\end{tabular}
\label{table:surf}
\end{table}

\begin{figure}
  \includegraphics[width=\linewidth]{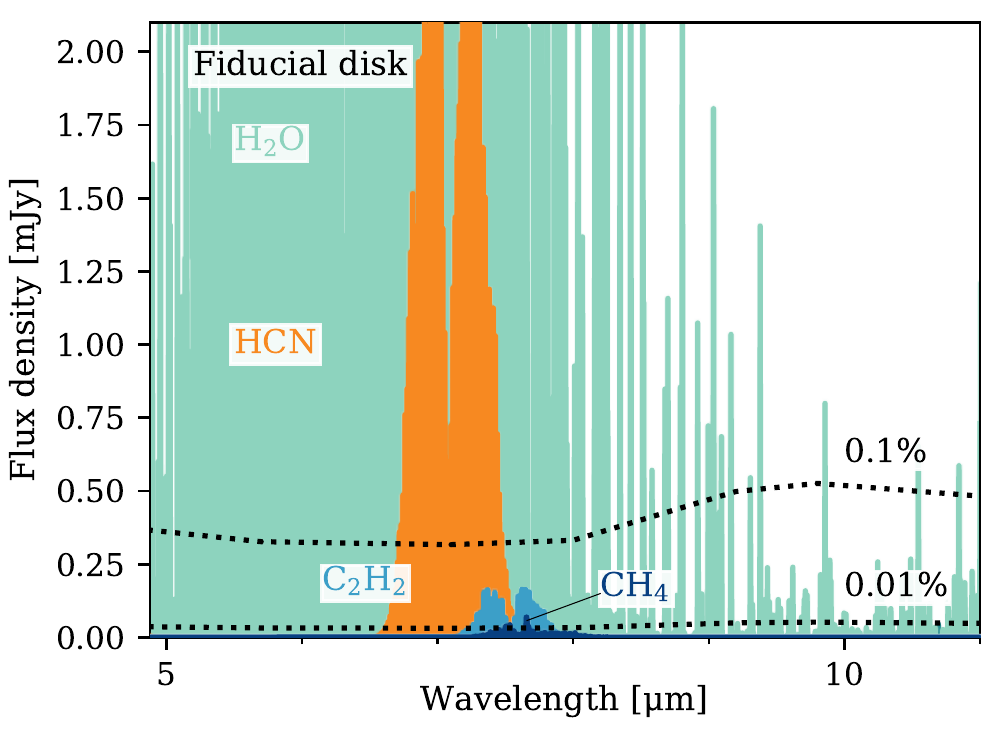}
  \includegraphics[width=\linewidth]{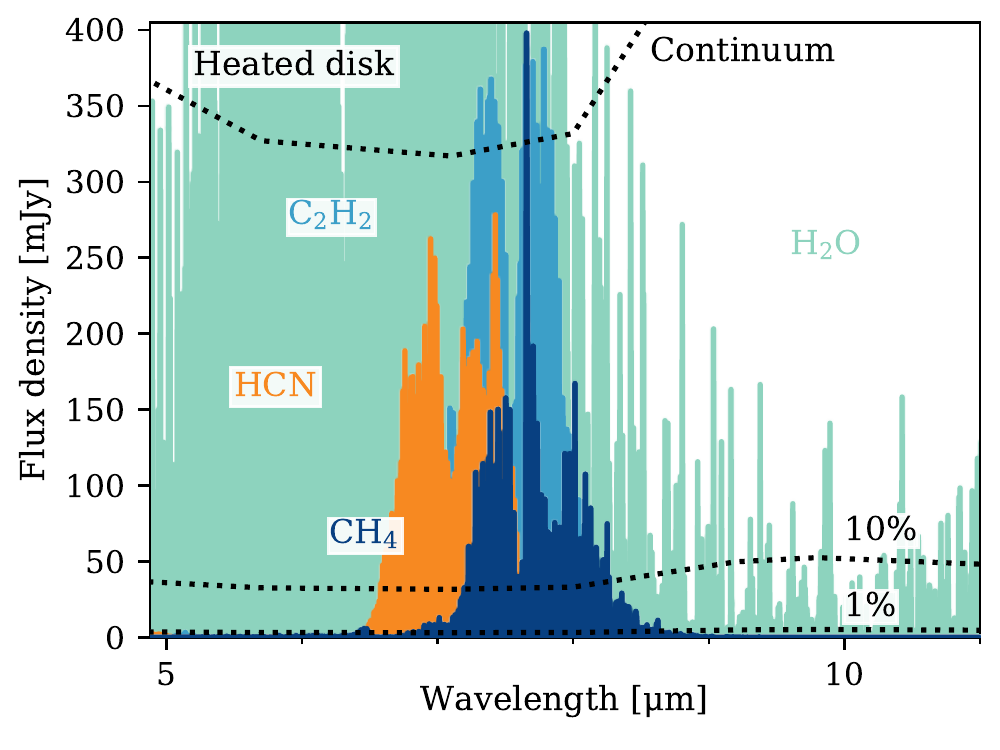}
  \caption{{ Zoom-in of spectra around the CH$_4$ feature for the fiducial model (top) and the model with artificially enhanced gas temperatures (bottom). Dotted lines indicate the level of the continuum flux density, assumed to equal the median spectral energy distribution (SED) for low mass stars in Taurus \citep{2006ApJS..165..568F}, and fractions thereof. A signal-to-noise ratio (SNR) $\gg$1000 would be required on the continuum to detect the CH$_4$ peak for the fiducial model, whereas the level of CH$_4$ emission for the heated disk is easily detectable.} }
  \label{fig:detect}
\end{figure}

\begin{figure}[t]
    \includegraphics[width=\linewidth, trim={0 0 0 0}, clip]{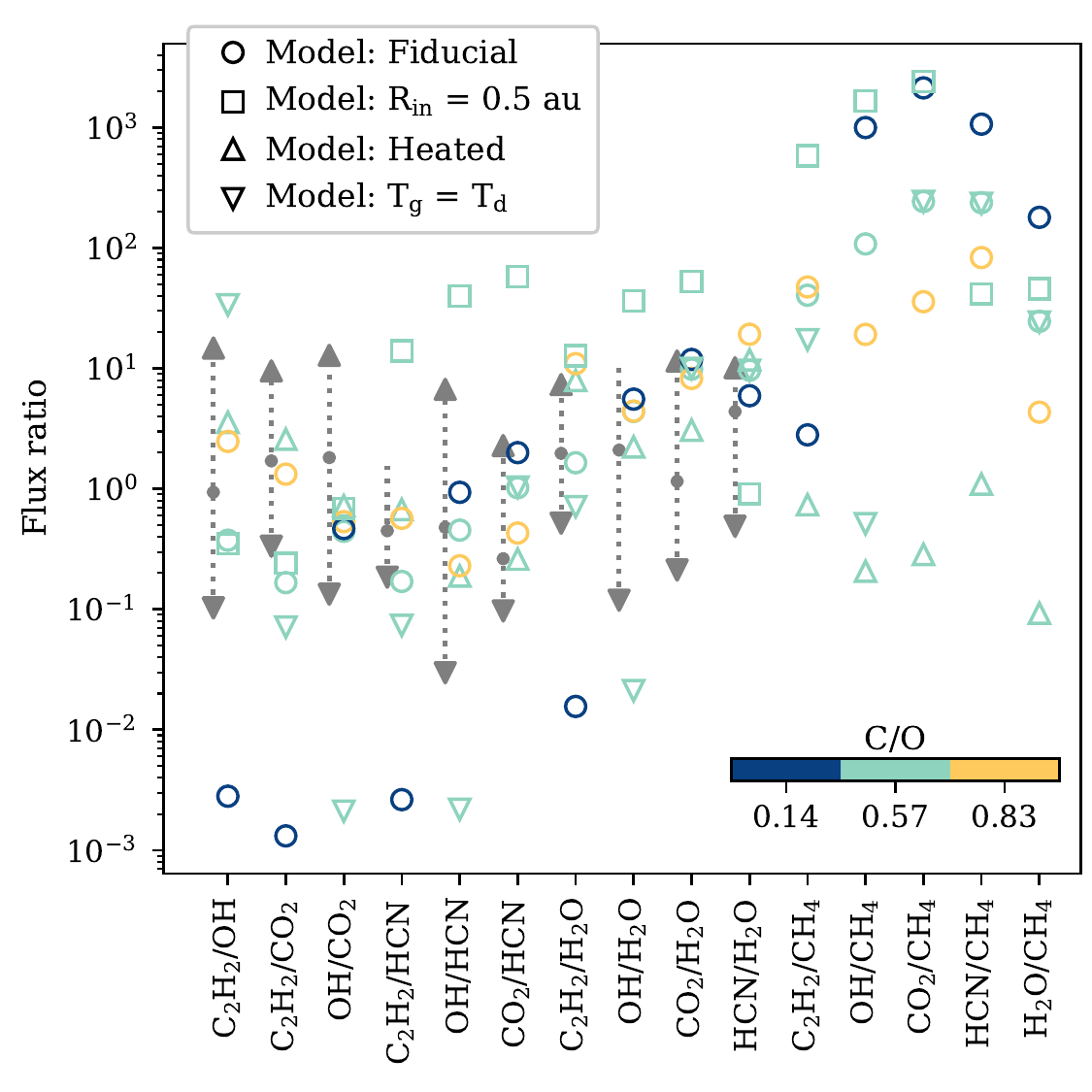}
    \caption{{Comparison of relative fluxes among C$_2$H$_2$, OH (23 $\mu$m), CO$_2$, HCN, H$_2$O, CH$_4$ for \textit{Spitzer} observations \citep{2011ApJ...731..130S} vs.~our modeled scenarios. Observed fluxes were included for disks where two or more spectral lines were detected.  The range of observed values or limits are shown, with arrows indicating where upper or lower limits were included. Gray dots represent the values for AA~Tau as an example for the flux ratios within a single source. Model results are provided in colors for C/O ratios and compositions as shown in Figure~\ref{fig:flux_C2O} and symbols indicating the following initial conditions: the fiducial physical disk model (circles), models where the  disk gas temperature was increased by a factor of 2 (upward triangles) or set equal to the dust temperature (downward triangles), and assuming an inner gas radius of 0.5~au instead of 0.2~au (squares).}} 
\label{fig:flux_ratios}
\end{figure}

Note that there are multiple major carbon carriers in the midplane for the fiducial disk. At 1~au, most of the total carbon is stored in CO, but carbon is also found in CO$_2$ at the 10\% level and both CH$_3$OH and CH$_4$ at single-digit percentages. In contrast, CO is by far the dominant carbon carrier for the disk model with an initial surface-like composition (Fig.~\ref{fig:surf2mid}). We find that even over 3$\times$10$^6$ years, the midplane abundances of CO$_2$, CH$_3$OH, and CH$_4$ in the model with an initial surface-like composition do not approach the levels seen in the fiducial disk initialized with interstellar ices. Therefore, if vigorous mixing is present, then it is possible that the entire column of material in the inner disk will tend toward the CO-dominated composition of the surface (see discussion in Section~\ref{section4.3}).

\section{Discussion}\label{section4}

\subsection{Predictions for \textit{JWST} MIRI}
In this section, we use our modeled fluxes for CO$_2$, C$_2$H$_2$, HCN,  H$_2$O, OH, and CH$_4$ to consider what will be possible with {\em JWST} MIRI. While MIRI is not the only {\em JWST} instrument that will be able to do molecular gas science in protoplanetary disks, the broad wavelength coverage and high sensitivity make it an excellent tool to survey the diversity of disk gas compositions. With the exception of CH$_4$, all of the aforementioned molecules have been observed in emission from T Tauri disks with \textit{Spitzer}. Therefore, {\em JWST} should be able to readily detect them and provide a statistical picture of disk chemistry. 

The potential to observe CH$_4$ is more unique to {\em JWST}, however we find it will be challenging. Particularly for gas temperatures similar to those in the fiducial disk model, we expect that CH$_4$ will be more difficult to detect than the other molecular species because of low absolute fluxes and low fluxes relative to other species with overlapping emission features. Figure~\ref{fig:detect} compares the modeled CH$_4$ emission to multiple line-to-continuum thresholds.  The CH$_4$ feature overlaps with emission from C$_2$H$_2$, HCN, and H$_2$O and is generally much weaker than that of the interfering molecular species. Only in the case of the artificially heated disk does CH$_4$ emission appear in the detectable range. Increasing the C/O ratio of the disk also helps by increasing the peak height of the CH$_4$ feature (see Appendix). 

Detected flux ratios among all molecular lines observed by \cite{2011ApJ...733..102C} and \cite{2011ApJ...731..130S} with \textit{Spitzer} generally fall in the range of 0.1 to 10. Overall, this is more consistent with models that have C/O ratios of 0.57--0.83, a full inner gas disk, and gas temperatures that are warmer than those of the dust (Figure~\ref{fig:flux_ratios}). In comparison, the C/O ratios furthest from the solar value result in much larger ranges in line flux ratios, spanning from 0.001 to 1000 in the case of C/O = 0.14 for example. Given the current observational data, it appears that the modeled disk with an inner gas radius of 0.5~au is inconsistent with the observed limits on the C$_2$H$_2$/HCN and OH/H$_2$O flux ratios. In addition, the low OH fluxes in the model where the gas and dust are thermally coupled cause the flux ratios relative to OH to fall outside the observed range. However, this scenario cannot be ruled out given the unconstrained limits. The improved sensitivity of \textit{JWST} will provide access to a larger range of ratios and further constraints. 

Many line flux ratios show sensitivity to the C/O ratio, but these trends could often be conflated with changes in gas temperature or the inner disk gas radius. The HCN/H$_2$O flux ratio is the most consistent in tracking the C/O ratio, showing a lack of sensitivity to the variations in gas temperature tested here (Fig.~\ref{fig:flux_ratios}). HCN/H$_2$O has previously been identified as a promising tracer of the C/O ratio in the inner regions of disks \citep{2011ApJ...733..102C}. Although this flux ratio is also sensitive to the inner disk gas radius in our models, the disparity between the C$_2$H$_2$, OH, and CO$_2$ fluxes and those of HCN, H$_2$O, and CH$_4$ separates changes to this parameter from those to the C/O ratio. Whereas the C/O ratio reflects the composition of the gas, indicators of the inner gas edge will depend on differences in the emitting regions of the molecules. The high CH$_4$ fluxes in the heated disk cause flux ratios of other molecules relative to CH$_4$ to be 0.1--1.0, generally lower than those in cooler disk models. CH$_4$ may therefore be an indicator of different temperature regimes.

\subsection{Uncertainties in the Predicted Molecular Column Densities and Fluxes}

The actual column of material that is sampled via observations at infrared wavelengths will depend on the opacity of the dust and contributing gas species in the disk. The assumed H$_2$ surface column density of 10$^{23}$~cm$^{-2}$ could vary by up to a factor of $\sim$10 for bounding scenarios of pure silicate grains or pure graphite grains or alternatively when assuming that the total disk dust mass is present in small ($\lesssim$1~$\mu$m) grains. Given the steep gas density gradient in the vertical direction and chemical transitions that depend on the vertical profiles of the disk temperature and radiation field, predicted molecular column densities are very sensitive to the assumed depth of the surface column probed by infrared observations. Probing deeper into the disk would mean higher molecular column densities and compositions that are influenced by that of the shielded midplane. Alternatively, if observations probe a more limited region closer to the disk surface, then molecular column densities would be lower and the timescales for converting refractory and volatile carbon carriers into CO would be $<$100--1000~yrs at 1~au.

In addition, the exclusion of non-LTE effects may be a limitation of this work. LTE models have been used to interpret \textit{Spitzer} observations of line emission from numerous disks \cite[e.g.,][]{2011ApJ...731..130S, 2011ApJ...733..102C}. Modeling of mid-infrared H$_2$O emission by \cite{2009ApJ...704.1471M} suggests that including the decoupling of gas and dust counteracts the effects of a non-LTE treatment. A simple approach ignoring both gas decoupling and non-LTE effects may therefore be sufficient for H$_2$O. Meanwhile, detailed studies focused on HCN and CO$_2$ have predicted differences in IR band fluxes by up to a factor of three when using LTE versus non-LTE models \citep{2015A&A...575A..94B,2017A&A...601A..36B}. Our predictions are unlikely to be as strongly affected because our modeled molecular abundances drop off sharply in the tenuous surface layers of the disk where non-LTE effects are most important \citep[see discussion by][]{2018A&A...618A..57W}. 

\subsection{Tracking C \& O Carriers in Disks}\label{section4.3}
The insensitivity of the surface composition to initial abundances allows for determination of the elemental C/O ratio in the inner disk based on molecular tracers. This ratio will be useful for indirectly tracking changes in the disk composition and for observational comparisons between disks and planetary atmospheres. Planet compositions are traditionally considered to be the product of disk midplane materials, but some contribution may also be derived from larger scale heights via vertical accretion \citep[e.g.,][]{2019Natur.574..378T, 2020A&A...635A..68C}. The relative importance of the surface chemistry will depend on the flow rate of material from the surface. Given that surface chemistry erases the assumed initial abundances in our models within the disk lifetime (e.g., Fig.~\ref{fig:mid}), it is possible that the relative abundances of \textit{molecular species} derived from infrared observations may not be representative of the composition of planet-forming regions in the disk midplane.

The degree to which separate molecular compositions can be maintained in the surface vs.~midplane will depend on the efficiency of vertical mixing. The diffusion timescale for a single scale height at 1~au is about 15~yrs based on the formula for the diffusion coefficient of gas from \citet{2004A&A...421.1075D} and turbulence characterized by a standard $\alpha$ value of 0.01. However, uncertainty in the diffusion coefficient by orders of magnitude could largely vary this timescale estimate. The competing chemical timescale, the timescale needed to reach a steady-state surface composition, decreases sharply with height above the midplane and depends on the strength of the local UV radiation field. If the bulk of midplane materials reach a height where the surface chemistry occurs much more rapidly than the mixing timescale, the entire vertical column may adopt a more surface-like, CO-dominated composition (e.g., Fig.~\ref{fig:surf2mid}).  

Whereas infrared observations find that gas in the inner regions of T Tauri disks is CO-rich \citep{2011ApJ...731..130S}, detections of the H$_2$ isotopologue hydrogen deuteride (HD) in three disks by the \textit{Herschel Space Observatory} combined with submillimeter observations of CO isotopologues from the Submillimeter Array (SMA) and the Atacama Large Millimeter/submillimeter Array (ALMA) resulted in estimated cold disk CO/H$_2$ ratios of up to 100$\times$ below the interstellar value of $\sim$10$^{-4}$ \citep{2013Natur.493..644B, 2013ApJ...776L..38F, 2016ApJ...831..167M}. The apparent lack of gas-phase CO in the outer disk is not an uncommon finding. Earlier suggestions of CO depletion were made based on weak CO relative to mm continuum emission for disks BP Tau \citep{2003A&A...402.1003D}, CQ Tau, and MWC 758 \citep{2008A&A...488..565C}. In addition, total gas mass estimates based on observed CO emission are found to be lower than those based on sub-mm/mm dust assuming a gas-to-dust mass ratio of 100 for numerous disks in recent ALMA surveys \citep[e.g.,][]{2017A&A...599A.113M}. Studies of multiple disk gas tracers have also found examples of disk compositions that are consistent with a sub-interstellar CO abundance \citep{2018ApJ...865..155C, 2019ApJ...881..127A}. 

Some loss of gas-phase CO is expected as a result of freezeout in cooler regions of the disk (T$\lesssim$ 20 K) and photodissociation of CO at regions with low UV extinction. But radial distributions of CO obtained from spatially resolved ALMA observations of its isotopologues in the TW Hya disk show significant depletion relative to interstellar values even interior to the CO snowline and near the disk midplane \citep{2016ApJ...823...91S, 2017NatAs...1E.130Z}. An alternative explanation for the lack of CO is that chemical processes occurring in protoplanetary disk environments result in the sequestration of CO into less volatile carbon species over time \citep{2014FaDi..168...61B, 2014ApJ...790...97F, 2015A&A...579A..82R, 2016ApJ...822...53Y, 2016A&A...595A..83E}. \citet{2018ApJ...856...85S} find that chemical reprocessing can account for up to an order of magnitude of CO depletion under the right disk conditions, but reaching the levels of depletion suggested by the aforementioned HD measurements requires additional non-chemical processes such as trapping of CO in cooler regions via vertical mixing or into large, km-sized, bodies forming in the disk \citep{2018ApJ...864...78K, 2019ApJ...883...98Z}. Potential products of the chemical reprocessing of CO ice in protoplanetary disks are CH$_3$OH, CO$_2$, and CH$_4$ ices \citep{2018ApJ...856...85S, 2018A&A...618A.182B}. Our modeling suggests that such products would be converted to CO in a fraction of the disk lifetime in the inner disk surface layers probed at infrared wavelengths. Therefore, if chemical conversion is responsible for the ``missing'' CO in the outer disk and this carbon is retained in other volatile forms, it should become detectable again in infrared observations.

The carbon carriers from the outer disk could be introduced into the inner disk via radial drift. Radial drift occurs when dust grains in the disk grow large enough to decouple from the gas and begin to drift radially inward as a result of the difference in velocity between the dust and the slower-moving, pressure-supported gas.  This transport of materials may influence the composition of the inner disk \citep[e.g.,][]{2018A&A...611A..80B,2020ApJ...891L..16Z,2020ApJ...903..124B}. Contributions from larger radii via radial drift will depend on the partitioning of materials between the gas and solid phases in the outer disk. An influx of H$_2$O-rich ices \citep[e.g.,][]{2006Icar..181..178C} could enhance H$_2$O abundances (and lower the C/O ratio) in the inner disk. Alternatively, H$_2$O could be locked up in large solid bodies reducing the amount of oxygen \citep[and raising the C/O ratio;][]{2013ApJ...766..134N}. Refractory carbon materials or carbon-rich ices could also be delivered from the outer disk. Signatures of inward radial drift may be preserved in the gas if resupply occurs on a timescale of $<$10$^{3}$--10$^{4}$~yrs based on the time needed for an initial excess of carbon in grains, CO$_2$, CH$_3$OH, or CH$_4$ to be depleted in the inner few au as seen in Fig.~\ref{fig:mid}. Even if direct signatures are erased, the C/O ratio could provide indirect evidence for dynamical processes. Combining such evidence from infrared observations with those at longer wavelengths (e.g., ALMA or the future Square Kilometer Array) will be important for determining the distribution of major carbon and oxygen carriers throughout the planet-forming regions of protoplanetary disks. 

\vspace{1cm}
\section{Conclusion}\label{section5}
We present models of different chemical compositions in the inner regions of protoplanetary disks and predict how they will affect the fluxes of a set of observable carbon and oxygen tracers at mid-infrared wavelengths. Our results will inform future investigations of protoplanetary disks by observatories such as \textit{JWST}. Here we summarize our conclusions based on these analyses.

Our models show that, in the absence of mass transport, the composition of the inner disk surface (for radii of up to a few au) is dominated by UV and X-ray photon-driven chemistry, which largely resets the initial chemical state on a timescale less than the characteristic bulk disk gas lifetime of a few million years. CO is the dominant carbon carrier in the inner-disk steady-state surface composition regardless of the initial distribution of carbon among major carbon-bearing disk species (i.e., CO, CO$_2$, CH$_3$OH, CH$_4$, and small refractory carbon grains). This conversion of less-volatile carbon species to CO could explain the apparent discrepancy between CO-poor gas observed in the outer regions of protoplanetary disks by ALMA relative to the predominately CO-rich gas observed at infrared wavelengths. In contrast, midplane compositions reflect the initial chemical abundances (when disk dynamics are excluded). The degree to which the original carbon carriers are converted to different chemical species in the midplane is largely driven by ion chemistry and therefore tied to the cosmic-ray ionization rate.

Actual inner disk compositions will depend on the scale and efficiency of mass transport processes in the disk. Efficient vertical transport of materials may result in the destruction of materials that would otherwise be preserved over the disk lifetime if left in the midplane. Because a gas of surface-like composition will not return to the average disk midplane composition in less than $\sim$10$^6$ years, this could potentially result in the entire vertical column appearing more surface-like in composition. Rapid resupply of icy and refractory carbon rich solids via radial drift on timescales of less than 10$^3$--10$^4$~yrs could also alter inner disk (radii $\sim$1~au) surface compositions. 

Infrared observations may not directly probe the relative abundances of specific major carbon and oxygen carriers in the shielded planet forming regions of protoplantary disks. However, the predicted fluxes of C$_2$H$_2$, HCN, CO$_2$, H$_2$O, OH, and CH$_4$ are sensitive to changes in the C/O ratio, inner gas radius, and gas temperature of the disk. In particular, the HCN/H$_2$O flux ratio appears to be a good tracer of the disk C/O ratio even in different gas temperature regimes. Furthermore, molecular flux ratios relative to CH$_4$ may serve as an indicator of the gas temperature. Although we only considered the use of total integrated fluxes here, additional constraints on the location and temperature of disk emission will be available via line shapes in spectrally resolved observations. These results are promising for future comparisons of C/O ratios with the atmospheres of extrasolar gas giant planets and investigating the physical conditions of the inner regions of protoplanetary disks. 

\software {Astropy \citep{astropy:2013, astropy:2018}, Matplotlib \citep{matplotlib}, Numpy \citep{numpy},  Pandas \citep{Pandas}, parallel \citep{tange_ole_2018_1146014}, slabspec \citep{slabspec_key}}

\acknowledgments The authors thank the anonymous reviewer for their helpful suggestions. D.E.A. acknowledges support from the Virginia Initiative on Cosmic Origins (VICO) Postdoctoral Fellowship. L.I.C. gratefully acknowledges support from the David and Lucille Packard Foundation, the Virginia Space Grant Consortium, Johnson \& Johnson's WiSTEM2D Award, and NSF AAG grant number AST-1910106. G.A.B. acknowledges support from the NASA XRP program (NNX16AB48G).

\bibliography{Anderson}

\begin{thebibliography}{}
\expandafter\ifx\csname natexlab\endcsname\relax\def\natexlab#1{#1}\fi
\providecommand{\url}[1]{\href{#1}{#1}}
\providecommand{\dodoi}[1]{doi:~\href{http://doi.org/#1}{\nolinkurl{#1}}}
\providecommand{\doeprint}[1]{\href{http://ascl.net/#1}{\nolinkurl{http://ascl.net/#1}}}
\providecommand{\doarXiv}[1]{\href{https://arxiv.org/abs/#1}{\nolinkurl{https://arxiv.org/abs/#1}}}

\bibitem[{{Ag{\'u}ndez} {et~al.}(2008){Ag{\'u}ndez}, {Cernicharo}, \&
  {Goicoechea}}]{2008A&A...483..831A}
{Ag{\'u}ndez}, M., {Cernicharo}, J., \& {Goicoechea}, J.~R. 2008, \aap, 483,
  831, \dodoi{10.1051/0004-6361:20077927}

\bibitem[{{Ag{\'u}ndez} {et~al.}(2018){Ag{\'u}ndez}, {Roueff}, {Le Petit}, \&
  {Le Bourlot}}]{2018A&A...616A..19A}
{Ag{\'u}ndez}, M., {Roueff}, E., {Le Petit}, F., \& {Le Bourlot}, J. 2018,
  \aap, 616, A19, \dodoi{10.1051/0004-6361/201732518}

\bibitem[{{Alata} {et~al.}(2014){Alata}, {Cruz-Diaz}, {Mu{\~n}oz Caro}, \&
  {Dartois}}]{2014A&A...569A.119A}
{Alata}, I., {Cruz-Diaz}, G.~A., {Mu{\~n}oz Caro}, G.~M., \& {Dartois}, E.
  2014, \aap, 569, A119, \dodoi{10.1051/0004-6361/201323118}

\bibitem[{{Alata} {et~al.}(2015){Alata}, {Jallat}, {Gavilan}, {Chabot},
  {Cruz-Diaz}, {Munoz Caro}, {B{\'e}roff}, \& {Dartois}}]{2015A&A...584A.123A}
{Alata}, I., {Jallat}, A., {Gavilan}, L., {et~al.} 2015, \aap, 584, A123,
  \dodoi{10.1051/0004-6361/201526368}

\bibitem[{{Anderson} {et~al.}(2017){Anderson}, {Bergin}, {Blake}, {Ciesla},
  {Visser}, \& {Lee}}]{2017ApJ...845...13A}
{Anderson}, D.~E., {Bergin}, E.~A., {Blake}, G.~A., {et~al.} 2017, \apj, 845,
  13, \dodoi{10.3847/1538-4357/aa7da1}

\bibitem[{{Anderson} {et~al.}(2019){Anderson}, {Blake}, {Bergin}, {Zhang},
  {Carpenter}, {Schwarz}, {Huang}, \& {{\"O}berg}}]{2019ApJ...881..127A}
{Anderson}, D.~E., {Blake}, G.~A., {Bergin}, E.~A., {et~al.} 2019, \apj, 881,
  127, \dodoi{10.3847/1538-4357/ab2cb5}

\bibitem[{{Andrews} {et~al.}(2011){Andrews}, {Wilner}, {Espaillat}, {Hughes},
  {Dullemond}, {McClure}, {Qi}, \& {Brown}}]{2011ApJ...732...42A}
{Andrews}, S.~M., {Wilner}, D.~J., {Espaillat}, C., {et~al.} 2011, \apj, 732,
  42, \dodoi{10.1088/0004-637X/732/1/42}

\bibitem[{{Astropy Collaboration} {et~al.}(2013){Astropy Collaboration},
  {Robitaille}, {Tollerud}, {Greenfield}, {Droettboom}, {Bray}, {Aldcroft},
  {Davis}, {Ginsburg}, {Price-Whelan}, {Kerzendorf}, {Conley}, {Crighton},
  {Barbary}, {Muna}, {Ferguson}, {Grollier}, {Parikh}, {Nair}, {Unther},
  {Deil}, {Woillez}, {Conseil}, {Kramer}, {Turner}, {Singer}, {Fox}, {Weaver},
  {Zabalza}, {Edwards}, {Azalee Bostroem}, {Burke}, {Casey}, {Crawford},
  {Dencheva}, {Ely}, {Jenness}, {Labrie}, {Lim}, {Pierfederici}, {Pontzen},
  {Ptak}, {Refsdal}, {Servillat}, \& {Streicher}}]{astropy:2013}
{Astropy Collaboration}, {Robitaille}, T.~P., {Tollerud}, E.~J., {et~al.} 2013,
  \aap, 558, A33, \dodoi{10.1051/0004-6361/201322068}

\bibitem[{{Banzatti} {et~al.}(2020){Banzatti}, {Pascucci}, {Bosman}, {Pinilla},
  {Salyk}, {Herczeg}, {Pontoppidan}, {Vazquez}, {Watkins}, {Krijt}, {Hendler},
  \& {Long}}]{2020ApJ...903..124B}
{Banzatti}, A., {Pascucci}, I., {Bosman}, A.~D., {et~al.} 2020, \apj, 903, 124,
  \dodoi{10.3847/1538-4357/abbc1a}

\bibitem[{{Bauer} {et~al.}(1997){Bauer}, {Finocchi}, {Duschl}, {Gail}, \&
  {Schloeder}}]{1997A&A...317..273B}
{Bauer}, I., {Finocchi}, F., {Duschl}, W.~J., {Gail}, H.-P., \& {Schloeder},
  J.~P. 1997, \aap, 317, 273

\bibitem[{{Bergin} {et~al.}(2015){Bergin}, {Blake}, {Ciesla}, {Hirschmann}, \&
  {Li}}]{2015PNAS..112.8965B}
{Bergin}, E.~A., {Blake}, G.~A., {Ciesla}, F., {Hirschmann}, M.~M., \& {Li}, J.
  2015, Proceedings of the National Academy of Science, 112, 8965,
  \dodoi{10.1073/pnas.1500954112}

\bibitem[{{Bergin} {et~al.}(2014){Bergin}, {Cleeves}, {Crockett}, \&
  {Blake}}]{2014FaDi..168...61B}
{Bergin}, E.~A., {Cleeves}, L.~I., {Crockett}, N., \& {Blake}, G.~A. 2014,
  Faraday Discussions, 168, \dodoi{10.1039/C4FD00003J}

\bibitem[{{Bergin} {et~al.}(2013){Bergin}, {Cleeves}, {Gorti}, {Zhang},
  {Blake}, {Green}, {Andrews}, {Evans}, {Henning}, {{\"O}berg}, {Pontoppidan},
  {Qi}, {Salyk}, \& {van Dishoeck}}]{2013Natur.493..644B}
{Bergin}, E.~A., {Cleeves}, L.~I., {Gorti}, U., {et~al.} 2013, \nat, 493, 644,
  \dodoi{10.1038/nature11805}

\bibitem[{{Bethell} \& {Bergin}(2011{\natexlab{a}})}]{2011ApJ...740....7B}
{Bethell}, T.~J., \& {Bergin}, E.~A. 2011{\natexlab{a}}, \apj, 740, 7,
  \dodoi{10.1088/0004-637X/740/1/7}

\bibitem[{{Bethell} \& {Bergin}(2011{\natexlab{b}})}]{2011ApJ...739...78B}
---. 2011{\natexlab{b}}, \apj, 739, 78, \dodoi{10.1088/0004-637X/739/2/78}

\bibitem[{{Boogert} {et~al.}(2015){Boogert}, {Gerakines}, \&
  {Whittet}}]{2015ARA&A..53..541B}
{Boogert}, A.~C.~A., {Gerakines}, P.~A., \& {Whittet}, D. C.~B. 2015, \araa,
  53, 541, \dodoi{10.1146/annurev-astro-082214-122348}

\bibitem[{{Bosman} {et~al.}(2017){Bosman}, {Bruderer}, \& {van
  Dishoeck}}]{2017A&A...601A..36B}
{Bosman}, A.~D., {Bruderer}, S., \& {van Dishoeck}, E.~F. 2017, \aap, 601, A36,
  \dodoi{10.1051/0004-6361/201629946}

\bibitem[{{Bosman} {et~al.}(2018{\natexlab{a}}){Bosman}, {Tielens}, \& {van
  Dishoeck}}]{2018A&A...611A..80B}
{Bosman}, A.~D., {Tielens}, A.~G.~G.~M., \& {van Dishoeck}, E.~F.
  2018{\natexlab{a}}, \aap, 611, A80, \dodoi{10.1051/0004-6361/201732056}

\bibitem[{{Bosman} {et~al.}(2018{\natexlab{b}}){Bosman}, {Walsh}, \& {van
  Dishoeck}}]{2018A&A...618A.182B}
{Bosman}, A.~D., {Walsh}, C., \& {van Dishoeck}, E.~F. 2018{\natexlab{b}},
  \aap, 618, A182, \dodoi{10.1051/0004-6361/201833497}

\bibitem[{{Bruderer}(2013)}]{2013A&A...559A..46B}
{Bruderer}, S. 2013, \aap, 559, A46, \dodoi{10.1051/0004-6361/201321171}

\bibitem[{{Bruderer} {et~al.}(2015){Bruderer}, {Harsono}, \& {van
  Dishoeck}}]{2015A&A...575A..94B}
{Bruderer}, S., {Harsono}, D., \& {van Dishoeck}, E.~F. 2015, \aap, 575, A94,
  \dodoi{10.1051/0004-6361/201425009}

\bibitem[{{Carr} \& {Najita}(2008)}]{2008Sci...319.1504C}
{Carr}, J.~S., \& {Najita}, J.~R. 2008, Science, 319, 1504,
  \dodoi{10.1126/science.1153807}

\bibitem[{{Carr} \& {Najita}(2011)}]{2011ApJ...733..102C}
---. 2011, \apj, 733, 102, \dodoi{10.1088/0004-637X/733/2/102}

\bibitem[{{Chapillon} {et~al.}(2008){Chapillon}, {Guilloteau}, {Dutrey}, \&
  {Pi{\'e}tu}}]{2008A&A...488..565C}
{Chapillon}, E., {Guilloteau}, S., {Dutrey}, A., \& {Pi{\'e}tu}, V. 2008, \aap,
  488, 565, \dodoi{10.1051/0004-6361:200809523}

\bibitem[{{Ciesla} \& {Cuzzi}(2006)}]{2006Icar..181..178C}
{Ciesla}, F.~J., \& {Cuzzi}, J.~N. 2006, \icarus, 181, 178,
  \dodoi{10.1016/j.icarus.2005.11.009}

\bibitem[{{Cleeves} {et~al.}(2014){Cleeves}, {Bergin}, \&
  {Adams}}]{2014ApJ...794..123C}
{Cleeves}, L.~I., {Bergin}, E.~A., \& {Adams}, F.~C. 2014, \apj, 794, 123,
  \dodoi{10.1088/0004-637X/794/2/123}

\bibitem[{{Cleeves} {et~al.}(2015){Cleeves}, {Bergin}, {Qi}, {Adams}, \&
  {{\"O}berg}}]{2015ApJ...799..204C}
{Cleeves}, L.~I., {Bergin}, E.~A., {Qi}, C., {Adams}, F.~C., \& {{\"O}berg},
  K.~I. 2015, \apj, 799, 204, \dodoi{10.1088/0004-637X/799/2/204}

\bibitem[{{Cleeves} {et~al.}(2018){Cleeves}, {{\"O}berg}, {Wilner}, {Huang},
  {Loomis}, {Andrews}, \& {Guzman}}]{2018ApJ...865..155C}
{Cleeves}, L.~I., {{\"O}berg}, K.~I., {Wilner}, D.~J., {et~al.} 2018, \apj,
  865, 155, \dodoi{10.3847/1538-4357/aade96}

\bibitem[{{Cridland} {et~al.}(2020){Cridland}, {Bosman}, \& {van
  Dishoeck}}]{2020A&A...635A..68C}
{Cridland}, A.~J., {Bosman}, A.~D., \& {van Dishoeck}, E.~F. 2020, \aap, 635,
  A68, \dodoi{10.1051/0004-6361/201936858}

\bibitem[{{Cridland} {et~al.}(2016){Cridland}, {Pudritz}, \&
  {Alessi}}]{2016MNRAS.461.3274C}
{Cridland}, A.~J., {Pudritz}, R.~E., \& {Alessi}, M. 2016, \mnras, 461, 3274,
  \dodoi{10.1093/mnras/stw1511}

\bibitem[{{Draine} \& {Lee}(1984)}]{1984ApJ...285...89D}
{Draine}, B.~T., \& {Lee}, H.~M. 1984, \apj, 285, 89, \dodoi{10.1086/162480}

\bibitem[{{Dullemond} \& {Dominik}(2004)}]{2004A&A...421.1075D}
{Dullemond}, C.~P., \& {Dominik}, C. 2004, \aap, 421, 1075,
  \dodoi{10.1051/0004-6361:20040284}

\bibitem[{{Dutrey} {et~al.}(2003){Dutrey}, {Guilloteau}, \&
  {Simon}}]{2003A&A...402.1003D}
{Dutrey}, A., {Guilloteau}, S., \& {Simon}, M. 2003, \aap, 402, 1003,
  \dodoi{10.1051/0004-6361:20030317}

\bibitem[{{Eistrup} {et~al.}(2016){Eistrup}, {Walsh}, \& {van
  Dishoeck}}]{2016A&A...595A..83E}
{Eistrup}, C., {Walsh}, C., \& {van Dishoeck}, E.~F. 2016, \aap, 595, A83,
  \dodoi{10.1051/0004-6361/201628509}

\bibitem[{{Eistrup} {et~al.}(2018){Eistrup}, {Walsh}, \& {van
  Dishoeck}}]{2018A&A...613A..14E}
---. 2018, \aap, 613, A14, \dodoi{10.1051/0004-6361/201731302}

\bibitem[{{Favre} {et~al.}(2013){Favre}, {Cleeves}, {Bergin}, {Qi}, \&
  {Blake}}]{2013ApJ...776L..38F}
{Favre}, C., {Cleeves}, L.~I., {Bergin}, E.~A., {Qi}, C., \& {Blake}, G.~A.
  2013, \apjl, 776, L38, \dodoi{10.1088/2041-8205/776/2/L38}

\bibitem[{{Finocchi} {et~al.}(1997){Finocchi}, {Gail}, \&
  {Duschl}}]{1997A&A...325.1264F}
{Finocchi}, F., {Gail}, H.-P., \& {Duschl}, W.~J. 1997, \aap, 325, 1264

\bibitem[{{Fogel} {et~al.}(2011){Fogel}, {Bethell}, {Bergin}, {Calvet}, \&
  {Semenov}}]{2011ApJ...726...29F}
{Fogel}, J.~K.~J., {Bethell}, T.~J., {Bergin}, E.~A., {Calvet}, N., \&
  {Semenov}, D. 2011, \apj, 726, 29, \dodoi{10.1088/0004-637X/726/1/29}

\bibitem[{{Furlan} {et~al.}(2006){Furlan}, {Hartmann}, {Calvet}, {D'Alessio},
  {Franco-Hern{\'a}ndez}, {Forrest}, {Watson}, {Uchida}, {Sargent}, {Green},
  {Keller}, \& {Herter}}]{2006ApJS..165..568F}
{Furlan}, E., {Hartmann}, L., {Calvet}, N., {et~al.} 2006, \apjs, 165, 568,
  \dodoi{10.1086/505468}

\bibitem[{{Furuya} \& {Aikawa}(2014)}]{2014ApJ...790...97F}
{Furuya}, K., \& {Aikawa}, Y. 2014, \apj, 790, 97,
  \dodoi{10.1088/0004-637X/790/2/97}

\bibitem[{{Gail}(2001)}]{2001A&A...378..192G}
{Gail}, H.-P. 2001, \aap, 378, 192, \dodoi{10.1051/0004-6361:20011130}

\bibitem[{{Gail}(2002)}]{2002A&A...390..253G}
---. 2002, \aap, 390, 253, \dodoi{10.1051/0004-6361:20020614}

\bibitem[{{Gail} \& {Trieloff}(2017)}]{2017A&A...606A..16G}
{Gail}, H.-P., \& {Trieloff}, M. 2017, \aap, 606, A16,
  \dodoi{10.1051/0004-6361/201730480}

\bibitem[{{Gibb} \& {Horne}(2013)}]{2013ApJ...776L..28G}
{Gibb}, E.~L., \& {Horne}, D. 2013, \apjl, 776, L28,
  \dodoi{10.1088/2041-8205/776/2/L28}

\bibitem[{{Gordon} {et~al.}(2017){Gordon}, {Rothman}, {Hill}, {Kochanov},
  {Tan}, {Bernath}, {Birk}, {Boudon}, {Campargue}, {Chance}, {Drouin}, {Flaud},
  {Gamache}, {Hodges}, {Jacquemart}, {Perevalov}, {Perrin}, {Shine}, {Smith},
  {Tennyson}, {Toon}, {Tran}, {Tyuterev}, {Barbe}, {Cs{\'a}sz{\'a}r}, {Devi},
  {Furtenbacher}, {Harrison}, {Hartmann}, {Jolly}, {Johnson}, {Karman},
  {Kleiner}, {Kyuberis}, {Loos}, {Lyulin}, {Massie}, {Mikhailenko},
  {Moazzen-Ahmadi}, {M{\"u}ller}, {Naumenko}, {Nikitin}, {Polyansky}, {Rey},
  {Rotger}, {Sharpe}, {Sung}, {Starikova}, {Tashkun}, {Auwera}, {Wagner},
  {Wilzewski}, {Wcis{\l}o}, {Yu}, \& {Zak}}]{2017JQSRT.203....3G}
{Gordon}, I.~E., {Rothman}, L.~S., {Hill}, C., {et~al.} 2017, \jqsrt, 203, 3,
  \dodoi{10.1016/j.jqsrt.2017.06.038}

\bibitem[{{Habing}(1968)}]{1968BAN....19..421H}
{Habing}, H.~J. 1968, \bain, 19, 421

\bibitem[{{Harada} {et~al.}(2010){Harada}, {Herbst}, \&
  {Wakelam}}]{2010ApJ...721.1570H}
{Harada}, N., {Herbst}, E., \& {Wakelam}, V. 2010, \apj, 721, 1570,
  \dodoi{10.1088/0004-637X/721/2/1570}

\bibitem[{{Harries} {et~al.}(2004){Harries}, {Monnier}, {Symington}, \&
  {Kurosawa}}]{2004MNRAS.350..565H}
{Harries}, T.~J., {Monnier}, J.~D., {Symington}, N.~H., \& {Kurosawa}, R. 2004,
  \mnras, 350, 565, \dodoi{10.1111/j.1365-2966.2004.07668.x}

\bibitem[{{Heays} {et~al.}(2017){Heays}, {Bosman}, \& {van
  Dishoeck}}]{2017A&A...602A.105H}
{Heays}, A.~N., {Bosman}, A.~D., \& {van Dishoeck}, E.~F. 2017, \aap, 602,
  A105, \dodoi{10.1051/0004-6361/201628742}

\bibitem[{{Hunter}(2007)}]{matplotlib}
{Hunter}, J.~D. 2007, Computing in Science and Engineering, 9, 90,
  \dodoi{10.1109/MCSE.2007.55}

\bibitem[{{Krijt} {et~al.}(2018){Krijt}, {Schwarz}, {Bergin}, \&
  {Ciesla}}]{2018ApJ...864...78K}
{Krijt}, S., {Schwarz}, K.~R., {Bergin}, E.~A., \& {Ciesla}, F.~J. 2018, \apj,
  864, 78, \dodoi{10.3847/1538-4357/aad69b}

\bibitem[{{Lee} {et~al.}(2010){Lee}, {Bergin}, \&
  {Nomura}}]{2010ApJ...710L..21L}
{Lee}, J.-E., {Bergin}, E.~A., \& {Nomura}, H. 2010, \apjl, 710, L21,
  \dodoi{10.1088/2041-8205/710/1/L21}

\bibitem[{{Lynden-Bell} \& {Pringle}(1974)}]{1974MNRAS.168..603L}
{Lynden-Bell}, D., \& {Pringle}, J.~E. 1974, \mnras, 168, 603,
  \dodoi{10.1093/mnras/168.3.603}

\bibitem[{{Madhusudhan}(2019)}]{2019ARA&A..57..617M}
{Madhusudhan}, N. 2019, \araa, 57, 617,
  \dodoi{10.1146/annurev-astro-081817-051846}

\bibitem[{{Mathis} {et~al.}(1977){Mathis}, {Rumpl}, \&
  {Nordsieck}}]{1977ApJ...217..425M}
{Mathis}, J.~S., {Rumpl}, W., \& {Nordsieck}, K.~H. 1977, \apj, 217, 425,
  \dodoi{10.1086/155591}

\bibitem[{{McClure} {et~al.}(2016){McClure}, {Bergin}, {Cleeves}, {van
  Dishoeck}, {Blake}, {Evans}, {Green}, {Henning}, {{\"O}berg}, {Pontoppidan},
  \& {Salyk}}]{2016ApJ...831..167M}
{McClure}, M.~K., {Bergin}, E.~A., {Cleeves}, L.~I., {et~al.} 2016, \apj, 831,
  167, \dodoi{10.3847/0004-637X/831/2/167}

\bibitem[{{McElroy} {et~al.}(2013){McElroy}, {Walsh}, {Markwick}, {Cordiner},
  {Smith}, \& {Millar}}]{2013A&A...550A..36M}
{McElroy}, D., {Walsh}, C., {Markwick}, A.~J., {et~al.} 2013, \aap, 550, A36,
  \dodoi{10.1051/0004-6361/201220465}

\bibitem[{McKinney(2010)}]{Pandas}
McKinney, W. 2010, in Proceedings of the 9th Python in Science Conference, ed.
  S.~van~der Walt \& J.~Millman, 51 -- 56

\bibitem[{{Meijerink} {et~al.}(2009){Meijerink}, {Pontoppidan}, {Blake},
  {Poelman}, \& {Dullemond}}]{2009ApJ...704.1471M}
{Meijerink}, R., {Pontoppidan}, K.~M., {Blake}, G.~A., {Poelman}, D.~R., \&
  {Dullemond}, C.~P. 2009, \apj, 704, 1471,
  \dodoi{10.1088/0004-637X/704/2/1471}

\bibitem[{{Miotello} {et~al.}(2017){Miotello}, {van Dishoeck}, {Williams},
  {Ansdell}, {Guidi}, {Hogerheijde}, {Manara}, {Tazzari}, {Testi}, {van der
  Marel}, \& {van Terwisga}}]{2017A&A...599A.113M}
{Miotello}, A., {van Dishoeck}, E.~F., {Williams}, J.~P., {et~al.} 2017, \aap,
  599, A113, \dodoi{10.1051/0004-6361/201629556}

\bibitem[{Mishra \& Li(2015)}]{Mishra_2015}
Mishra, A., \& Li, A. 2015, The Astrophysical Journal, 809, 120,
  \dodoi{10.1088/0004-637x/809/2/120}

\bibitem[{{Najita} {et~al.}(2011){Najita}, {{\'A}d{\'a}mkovics}, \&
  {Glassgold}}]{2011ApJ...743..147N}
{Najita}, J.~R., {{\'A}d{\'a}mkovics}, M., \& {Glassgold}, A.~E. 2011, \apj,
  743, 147, \dodoi{10.1088/0004-637X/743/2/147}

\bibitem[{{Najita} {et~al.}(2013){Najita}, {Carr}, {Pontoppidan}, {Salyk}, {van
  Dishoeck}, \& {Blake}}]{2013ApJ...766..134N}
{Najita}, J.~R., {Carr}, J.~S., {Pontoppidan}, K.~M., {et~al.} 2013, \apj, 766,
  134, \dodoi{10.1088/0004-637X/766/2/134}

\bibitem[{{{\"O}berg} {et~al.}(2011){{\"O}berg}, {Murray-Clay}, \&
  {Bergin}}]{2011ApJ...743L..16O}
{{\"O}berg}, K.~I., {Murray-Clay}, R., \& {Bergin}, E.~A. 2011, \apjl, 743,
  L16, \dodoi{10.1088/2041-8205/743/1/L16}

\bibitem[{{Pascucci} {et~al.}(2013){Pascucci}, {Herczeg}, {Carr}, \&
  {Bruderer}}]{2013ApJ...779..178P}
{Pascucci}, I., {Herczeg}, G., {Carr}, J.~S., \& {Bruderer}, S. 2013, \apj,
  779, 178, \dodoi{10.1088/0004-637X/779/2/178}

\bibitem[{{Pontoppidan} {et~al.}(2010){Pontoppidan}, {Salyk}, {Blake},
  {Meijerink}, {Carr}, \& {Najita}}]{2010ApJ...720..887P}
{Pontoppidan}, K.~M., {Salyk}, C., {Blake}, G.~A., {et~al.} 2010, \apj, 720,
  887, \dodoi{10.1088/0004-637X/720/1/887}

\bibitem[{{Price} {et~al.}(2020){Price}, {Cleeves}, \&
  {{\"O}berg}}]{2020ApJ...890..154P}
{Price}, E.~M., {Cleeves}, L.~I., \& {{\"O}berg}, K.~I. 2020, \apj, 890, 154,
  \dodoi{10.3847/1538-4357/ab5fd4}

\bibitem[{{Price-Whelan} {et~al.}(2018){Price-Whelan}, {Sip{\H{o}}cz},
  {G{\"u}nther}, {Lim}, {Crawford}, {Conseil}, {Shupe}, {Craig}, {Dencheva},
  {Ginsburg}, {VanderPlas}, {Bradley}, {P{\'e}rez-Su{\'a}rez}, {de Val-Borro},
  {Paper Contributors}, {Aldcroft}, {Cruz}, {Robitaille}, {Tollerud},
  {Coordination Committee}, {Ardelean}, {Babej}, {Bach}, {Bachetti}, {Bakanov},
  {Bamford}, {Barentsen}, {Barmby}, {Baumbach}, {Berry}, {Biscani}, {Boquien},
  {Bostroem}, {Bouma}, {Brammer}, {Bray}, {Breytenbach}, {Buddelmeijer},
  {Burke}, {Calderone}, {Cano Rodr{\'\i}guez}, {Cara}, {Cardoso}, {Cheedella},
  {Copin}, {Corrales}, {Crichton}, {D{\textquoteright}Avella}, {Deil},
  {Depagne}, {Dietrich}, {Donath}, {Droettboom}, {Earl}, {Erben}, {Fabbro},
  {Ferreira}, {Finethy}, {Fox}, {Garrison}, {Gibbons}, {Goldstein}, {Gommers},
  {Greco}, {Greenfield}, {Groener}, {Grollier}, {Hagen}, {Hirst}, {Homeier},
  {Horton}, {Hosseinzadeh}, {Hu}, {Hunkeler}, {Ivezi{\'c}}, {Jain}, {Jenness},
  {Kanarek}, {Kendrew}, {Kern}, {Kerzendorf}, {Khvalko}, {King}, {Kirkby},
  {Kulkarni}, {Kumar}, {Lee}, {Lenz}, {Littlefair}, {Ma}, {Macleod},
  {Mastropietro}, {McCully}, {Montagnac}, {Morris}, {Mueller}, {Mumford},
  {Muna}, {Murphy}, {Nelson}, {Nguyen}, {Ninan}, {N{\"o}the}, {Ogaz}, {Oh},
  {Parejko}, {Parley}, {Pascual}, {Patil}, {Patil}, {Plunkett}, {Prochaska},
  {Rastogi}, {Reddy Janga}, {Sabater}, {Sakurikar}, {Seifert}, {Sherbert},
  {Sherwood-Taylor}, {Shih}, {Sick}, {Silbiger}, {Singanamalla}, {Singer},
  {Sladen}, {Sooley}, {Sornarajah}, {Streicher}, {Teuben}, {Thomas},
  {Tremblay}, {Turner}, {Terr{\'o}n}, {van Kerkwijk}, {de la Vega}, {Watkins},
  {Weaver}, {Whitmore}, {Woillez}, {Zabalza}, \& {Contributors}}]{astropy:2018}
{Price-Whelan}, A.~M., {Sip{\H{o}}cz}, B.~M., {G{\"u}nther}, H.~M., {et~al.}
  2018, \aj, 156, 123, \dodoi{10.3847/1538-3881/aabc4f}

\bibitem[{{Reboussin} {et~al.}(2015){Reboussin}, {Wakelam}, {Guilloteau},
  {Hersant}, \& {Dutrey}}]{2015A&A...579A..82R}
{Reboussin}, L., {Wakelam}, V., {Guilloteau}, S., {Hersant}, F., \& {Dutrey},
  A. 2015, \aap, 579, A82, \dodoi{10.1051/0004-6361/201525885}

\bibitem[{Salyk(2020)}]{slabspec_key}
Salyk, C. 2020, slabspec, v1.0,  Zenodo, \dodoi{10.5281/zenodo.4037306}.
\newblock \url{https://doi.org/10.5281/zenodo.4037306}

\bibitem[{{Salyk} {et~al.}(2011){Salyk}, {Pontoppidan}, {Blake}, {Najita}, \&
  {Carr}}]{2011ApJ...731..130S}
{Salyk}, C., {Pontoppidan}, K.~M., {Blake}, G.~A., {Najita}, J.~R., \& {Carr},
  J.~S. 2011, \apj, 731, 130, \dodoi{10.1088/0004-637X/731/2/130}

\bibitem[{{Savage} \& {Sembach}(1996)}]{1996ARA&A..34..279S}
{Savage}, B.~D., \& {Sembach}, K.~R. 1996, \araa, 34, 279,
  \dodoi{10.1146/annurev.astro.34.1.279}

\bibitem[{{Schwarz} {et~al.}(2016){Schwarz}, {Bergin}, {Cleeves}, {Blake},
  {Zhang}, {{\"O}berg}, {van Dishoeck}, \& {Qi}}]{2016ApJ...823...91S}
{Schwarz}, K.~R., {Bergin}, E.~A., {Cleeves}, L.~I., {et~al.} 2016, \apj, 823,
  91, \dodoi{10.3847/0004-637X/823/2/91}

\bibitem[{{Schwarz} {et~al.}(2018){Schwarz}, {Bergin}, {Cleeves}, {Zhang},
  {{\"O}berg}, {Blake}, \& {Anderson}}]{2018ApJ...856...85S}
---. 2018, \apj, 856, 85, \dodoi{10.3847/1538-4357/aaae08}

\bibitem[{Tange(2018)}]{tange_ole_2018_1146014}
Tange, O. 2018, GNU Parallel 2018 (Ole Tange), \dodoi{10.5281/zenodo.1146014}.
\newblock \url{https://doi.org/10.5281/zenodo.1146014}

\bibitem[{{Teague} {et~al.}(2019){Teague}, {Bae}, \&
  {Bergin}}]{2019Natur.574..378T}
{Teague}, R., {Bae}, J., \& {Bergin}, E.~A. 2019, \nat, 574, 378,
  \dodoi{10.1038/s41586-019-1642-0}

\bibitem[{{van der Walt} {et~al.}(2011){van der Walt}, {Colbert}, \&
  {Varoquaux}}]{numpy}
{van der Walt}, S., {Colbert}, S.~C., \& {Varoquaux}, G. 2011, Computing in
  Science and Engineering, 13, 22, \dodoi{10.1109/MCSE.2011.37}

\bibitem[{{van 't Hoff} {et~al.}(2020){van 't Hoff}, {Bergin}, {J{\o}rgensen},
  \& {Blake}}]{2020ApJ...897L..38V}
{van 't Hoff}, M. L.~R., {Bergin}, E.~A., {J{\o}rgensen}, J.~K., \& {Blake},
  G.~A. 2020, \apjl, 897, L38, \dodoi{10.3847/2041-8213/ab9f97}

\bibitem[{{Walsh} {et~al.}(2015){Walsh}, {Nomura}, \& {van
  Dishoeck}}]{2015A&A...582A..88W}
{Walsh}, C., {Nomura}, H., \& {van Dishoeck}, E. 2015, \aap, 582, A88,
  \dodoi{10.1051/0004-6361/201526751}

\bibitem[{{Wei} {et~al.}(2019){Wei}, {Nomura}, {Lee}, {Ip}, {Walsh}, \&
  {Millar}}]{2019ApJ...870..129W}
{Wei}, C.-E., {Nomura}, H., {Lee}, J.-E., {et~al.} 2019, \apj, 870, 129,
  \dodoi{10.3847/1538-4357/aaf390}

\bibitem[{{Wells} {et~al.}(2015){Wells}, {Pel}, {Glasse}, {Wright},
  {Aitink-Kroes}, {Azzollini}, {Beard}, {Brandl}, {Gallie}, {Geers}, {Glauser},
  {Hastings}, {Henning}, {Jager}, {Justtanont}, {Kruizinga}, {Lahuis}, {Lee},
  {Martinez-Delgado}, {Mart{\'\i}nez-Galarza}, {Meijers}, {Morrison},
  {M{\"u}ller}, {Nakos}, {O'Sullivan}, {Oudenhuysen}, {Parr-Burman}, {Pauwels},
  {Rohloff}, {Schmalzl}, {Sykes}, {Thelen}, {van Dishoeck}, {Vandenbussche},
  {Venema}, {Visser}, {Waters}, \& {Wright}}]{2015PASP..127..646W}
{Wells}, M., {Pel}, J.~W., {Glasse}, A., {et~al.} 2015, \pasp, 127, 646,
  \dodoi{10.1086/682281}

\bibitem[{{Woitke} {et~al.}(2018){Woitke}, {Min}, {Thi}, {Roberts}, {Carmona},
  {Kamp}, {M{\'e}nard}, \& {Pinte}}]{2018A&A...618A..57W}
{Woitke}, P., {Min}, M., {Thi}, W.~F., {et~al.} 2018, \aap, 618, A57,
  \dodoi{10.1051/0004-6361/201731460}

\bibitem[{{Yu} {et~al.}(2016){Yu}, {Willacy}, {Dodson-Robinson}, {Turner}, \&
  {Evans}}]{2016ApJ...822...53Y}
{Yu}, M., {Willacy}, K., {Dodson-Robinson}, S.~E., {Turner}, N.~J., \& {Evans},
  II, N.~J. 2016, \apj, 822, 53, \dodoi{10.3847/0004-637X/822/1/53}

\bibitem[{{Zhang} {et~al.}(2017){Zhang}, {Bergin}, {Blake}, {Cleeves}, \&
  {Schwarz}}]{2017NatAs...1E.130Z}
{Zhang}, K., {Bergin}, E.~A., {Blake}, G.~A., {Cleeves}, L.~I., \& {Schwarz},
  K.~R. 2017, Nature Astronomy, 1, 0130, \dodoi{10.1038/s41550-017-0130}

\bibitem[{{Zhang} {et~al.}(2019){Zhang}, {Bergin}, {Schwarz}, {Krijt}, \&
  {Ciesla}}]{2019ApJ...883...98Z}
{Zhang}, K., {Bergin}, E.~A., {Schwarz}, K., {Krijt}, S., \& {Ciesla}, F. 2019,
  \apj, 883, 98, \dodoi{10.3847/1538-4357/ab38b9}

\bibitem[{{Zhang} {et~al.}(2020){Zhang}, {Bosman}, \&
  {Bergin}}]{2020ApJ...891L..16Z}
{Zhang}, K., {Bosman}, A.~D., \& {Bergin}, E.~A. 2020, \apjl, 891, L16,
  \dodoi{10.3847/2041-8213/ab77ca}

\end{thebibliography}

 \appendix

\section{Updates to Reaction Network}
 
  Additional cosmic-ray, positive ion-neutral, neutral-neutral, electronic dissociative recombination, and photodissociation reactions were added from the OSU high-temperature network of \cite{2010ApJ...721.1570H}. As a result 11 new species were added, including carbon chains with 10 C atoms, C$_2$H$_6$, C$_2$H$_7^+$, and HN$_2$O$^+$. In addition, for select positive ion-neutral and ion-neutral radiative association reactions the temperature dependence of the reaction rate was altered to provide appropriate rates above 300 K \citep[see][]{2010ApJ...721.1570H}. Rates for UV photolysis induced by X-rays were also added for additional species based on efficiency values for cosmic-ray induced photoreactions from the UMIST database \citep{2013A&A...550A..36M}.

\begin{figure*}[h]
\includegraphics[width=\linewidth]{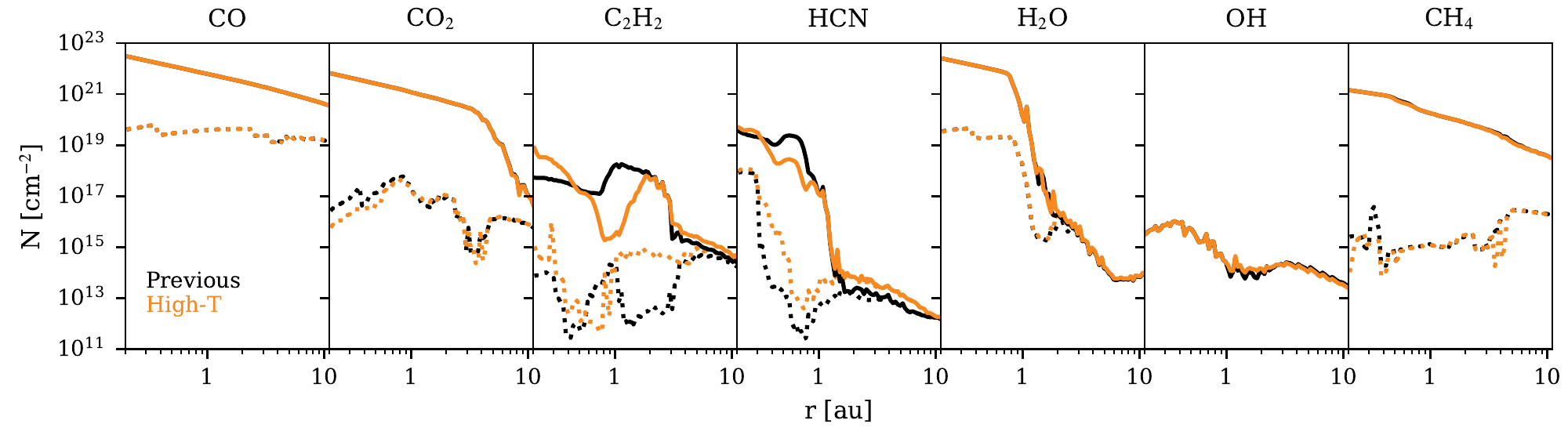}
    \caption{{Comparison of the surface column densities (dotted lines) and total column densities (from the surface to the midplane, solid lines) vs.~disk radius based on the previous reaction network (shown in black) to those based on the updated network incorporating high-temperature reactions (shown in orange). }} 
\label{fig:highT}
\end{figure*}

  Of the 653 chemical species modeled, about 80\% experienced more than 30\% change in their column densities for at least three consecutive radial grid points from 0.2--10 au. This is true when considering either the surface column density (calculated as described in Section \ref{section2.3}) or the total column density (calculated from the disk surface to the midplane). The percentage of species decreases to 67\% for changes by greater than a factor of 10. Chemical species containing multiple carbon atoms were among the most affected. Figure~\ref{fig:highT} compares the modeled column densities for the main species investigated in this work. The largest changes are seen for C$_2$H$_2$ and HCN.
  
 \section{Updates to Photodissociation Cross Sections}
 
 Updated photodissociation cross section data are from \cite{2017A&A...602A.105H}. The continuum was interpolated across the range of 912 to 2000 {\AA} with wavelength intervals of 10 {\AA}. Narrow lines were added to the appropriate wavelength bins. Photodissociation rates were calculated based on the UV field and cross section as described in \cite{2011ApJ...726...29F}. 

We added wavelength-dependent photodissociation cross sections for C$_2$, C$_2$H$_3$N, C$_2$H$_4$,  C$_2$H$_4$O, C$_2$H$_5$OH, C$_3$H, C$_3$H$_2$, C$_4$, C$_4$H,  C$_5$H, CH$_2$, CH$_3$, CH$_3$OH, CH$_5$N, CO$_2$, CS, H$_2^+$, H$_2$C$_3$, H$_2$CO,  H$_2$O$_2$, H$_2$S, H$_3^+$, HC$_3$, HC$_3$N, HCO, HCl, HF, N$_2$O, NH$_2$CHO, NO$_2$, OCS, OH$^+$, PH, SO, SiH, SiH$^+$, and SiO. The value for the linear isomer is used except in the cases of C$_3$H$_2$ (cyclic) \& H$_2$C$_3$ (linear) and C$_3$H (cyclic) \& HC$_3$ (linear) where both isomers are present in the network. We updated the wavelength-dependent photodissociation cross section data for C$_2$H$_2$, C$_3$, CH$_4$, CN, H$_2$O, HCN (and the same was used for HNC), NH, NH$_2$, NH$_3$, NO, O$_2$, and SO$_2$. The remaining species for which photodissociation rates are calculated based on wavelength-dependent rates from the literature include C$_2$H, C$_4$H$_2$, CH, CH$^+$, and OH. The treatment of H$_2$, CO, and N$_2$ self-shielding remains the same relative to previous versions of the model \citep[see][]{2011ApJ...726...29F,2018ApJ...865..155C}. 

\begin{figure*}[h]
\includegraphics[width=\linewidth]{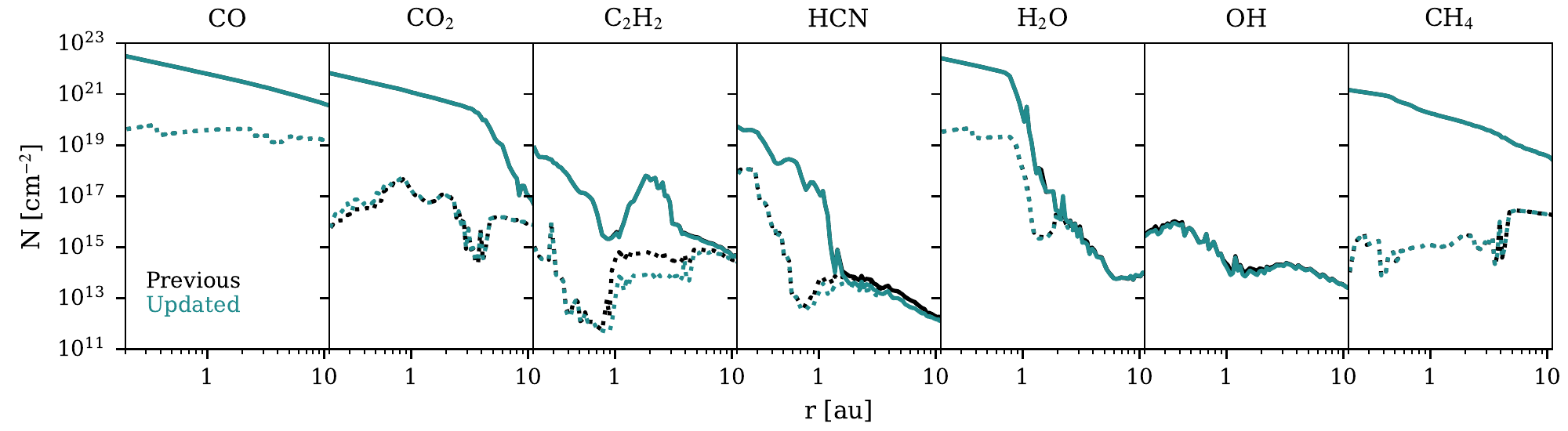}
    \caption{{Comparison of the surface column densities (dotted lines) and total column densities (from the surface to the midplane, solid lines) vs.~disk radius when using the previous photodissociation cross section data (shown in black) to those modeled when using the updated data (shown in blue-green). }} 
\label{fig:xsect}
\end{figure*}

Updating the photodissociation cross sections for the above listed species generally had a larger effect on the surface column densities than on the total column densities for modeled species. Of the 653 chemical species modeled, about 70\% experienced a change by more than 30\% in their surface column densities for at least three consecutive radial grid points from 0.2--10 au whereas only about 50\% experienced this level of change in their total column densities (calculated from the disk surface to the midplane). For changes over an order of magnitude these percentages are 30\% and 20\%, respectively. In many cases the amount of change was radially dependent. Among the species for which cross section data were updated in the model, the largest changes to surface column densities were for carbon-bearing species, in particular hydrocarbons, at select radial regions interior to 5 au. For the main species investigated in this work, differences in computed surface column densities due to the updated cross sections were less than a factor of 2--3 and limited to the inner 5 au from the central star. C$_2$H$_2$ was the exception, having over an order of magnitude decrease in the surface column density in the region around a few au (Figure~\ref{fig:xsect}). 

\section{Observability of Methane}
\begin{figure}
  \includegraphics[width=0.5\linewidth]{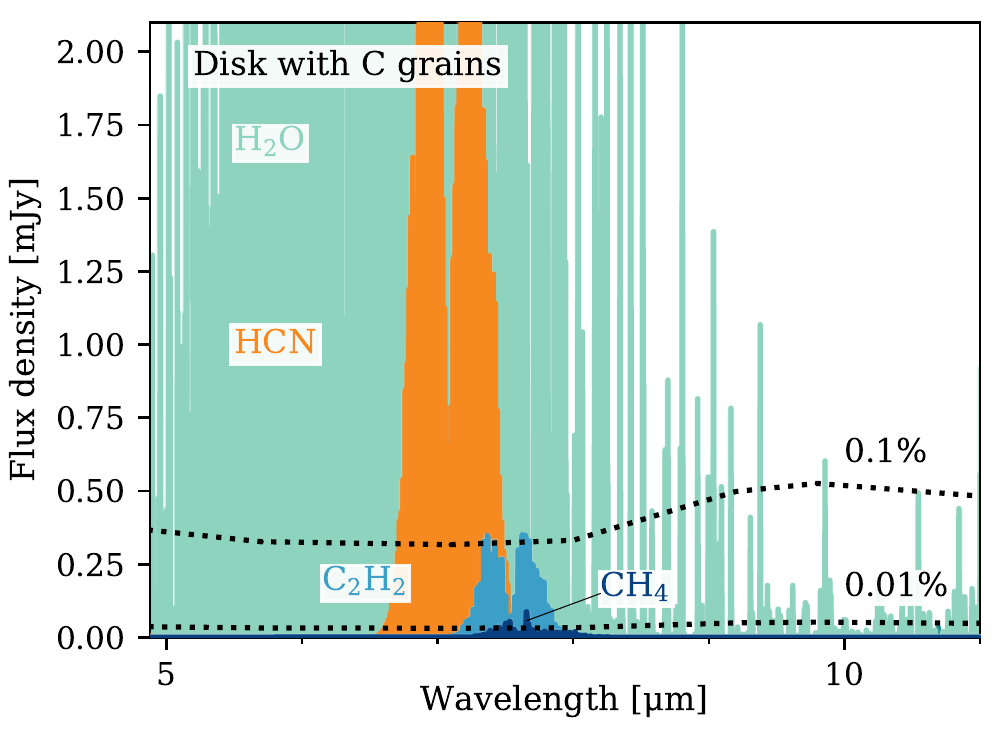}
  \includegraphics[width=0.5\linewidth]{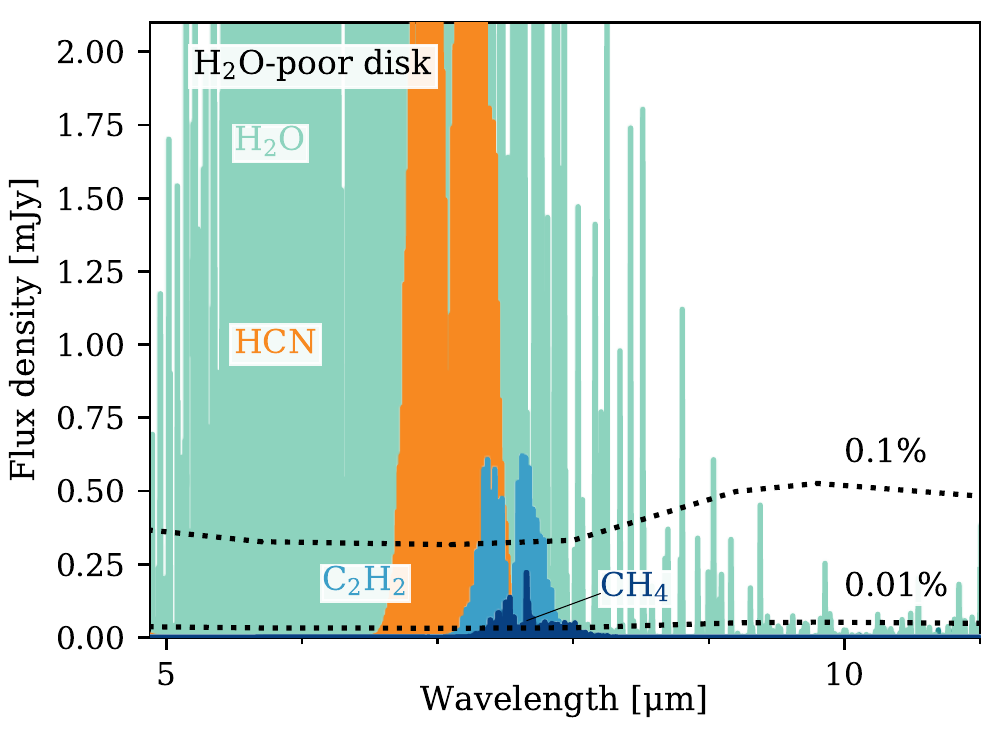}
  \caption{{Zoom-in of spectra around the CH$_4$ feature for the model including destruction of refractory carbon grains (left) and the model with initial H$_2$O abundances decreased by 10$\times$ (right). Dotted lines indicate fractions of the continuum flux density, assumed to equal the median spectral energy distribution (SED) for low mass stars in Taurus \citep{2006ApJS..165..568F}.}}

  \label{fig:detect2}
\end{figure}

Figure~\ref{fig:detect2} is a continuation of Figure~\ref{fig:detect} including additional models with different elemental C/O ratios.

\end{document}